\documentclass[a4paper,11pt]{article}
\pdfoutput=1 

\usepackage{jheppub} 

\usepackage[T1]{fontenc} 

\usepackage{amsmath,bm}
\usepackage{amsthm}
\usepackage{amsfonts}
\usepackage{amssymb}
\usepackage{graphicx}
\usepackage{float}
\usepackage{wrapfig}
\usepackage{latexsym}
\usepackage{hyperref}
\usepackage{feynmf}
\usepackage{exscale}
\usepackage{relsize}
\usepackage{listings}
\usepackage{upquote}
\usepackage{overpic}
\usepackage{setspace}
\usepackage{caption}
\usepackage{multirow}
\usepackage{mathtools}
\usepackage[utf8]{inputenc}
\usepackage{tikz}
\usetikzlibrary{calc,intersections,through,backgrounds}
\usepackage{comment}
\usepackage{url}
\usepackage{tensor}
\usepackage{subfig}

\newtheorem*{stat*}{Claims}
\newtheorem*{prop*}{Proposition}
\newtheorem*{conj*}{Conjecture}
\newtheorem*{theo*}{Theorem}

\def\be{\begin{equation}}
\def\ee{\end{equation}}
\def\ba{\begin{eqnarray}}
\def\ea{\end{eqnarray}}
\def\ban{\begin{align}}
\def\ean{\end{align}}

\def\b{\beta}

\def\b#1{\overline{#1}}

\def\CP1{\mathbb{CP}^1}
\def\SL2C{\mathrm{SL}(2,\mathbb{C})}

\def\Z2{\mathbb{Z}_2}

\def\<{\langle}
\def\>{\rangle}
\def\s{\sigma}

\usepackage{subfig}
\usetikzlibrary{arrows,decorations.pathmorphing,backgrounds,positioning,fit,shapes,chains,shapes.gates.logic.US,trees,tikzmark,decorations.text,calc,
mindmap,decorations.pathreplacing,decorations.markings}

\title{Scattering Forms, Worldsheet Forms and Amplitudes from Subspaces}

\author[a,b]{Song He}
\author[c,a]{,\,Gongwang Yan}
\author[a,b]{,\,Chi Zhang}
\author[a,b]{,\,Yong Zhang}
\affiliation[a]{CAS Key Laboratory of Theoretical Physics, Institute of Theoretical Physics, Chinese Academy
of Sciences, Beijing, 100190, China}
\affiliation[b]{
University of Chinese Academy of Sciences, No.19A Yuquan Road, Beijing 100049, China}
\affiliation[c]{Institute for Advanced Study, Tsinghua University, Beijing, 100084, China}

\emailAdd{songhe@itp.ac.cn}
\emailAdd{ygw17@mails.tsinghua.edu.cn}
\emailAdd{zhangchi@itp.ac.cn}
\emailAdd{yongzhang@itp.ac.cn}

\abstract{We present a general construction of two types of differential forms, based on any $(n{-}3)$-dimensional subspace in the kinematic space of $n$ massless particles. The first type is the so-called projective, scattering forms in kinematic space, while the second is defined in the moduli space of $n$-punctured Riemann spheres which we call worldsheet forms. We show that the pushforward of worldsheet forms, by summing over solutions of scattering equations, gives the corresponding scattering forms, which generalizes the results of~\cite{Arkani-Hamed:2017mur}. The pullback of scattering forms to subspaces can have natural interpretations as amplitudes in terms of Bern-Carrasco-Johansson double-copy construction or Cachazo-He-Yuan formula. As an application of our formalism, we construct in this way a large class of $d\log$ scattering forms and worldsheet forms, which are in one-to-one correspondence with non-planar MHV leading singularities in ${\cal N}=4$ super-Yang-Mills.  For every leading singularity function, we present a new determinant formula in moduli space, as well as a (combinatoric) polytope and associated scattering form in kinematic space. These include the so-called Cayley cases, where in each case the scattering form is the canonical forms of a convex polytope in the subspace, and scattering equations admit elegant rewritings as a map from the moduli space to the subspace.}

\begin{document}

 \tikzset{
particle/.style={draw=black, postaction={decorate}
}
 }

\maketitle

\section{Introduction}

A new framework has been proposed in~\cite{Arkani-Hamed:2017mur}, which naturally merges three lines of development in the study of new structures of scattering amplitudes: remarkable geometric constructions known as the amplituhedron in planar ${\cal N}=4$ SYM~\cite{Arkani-Hamed:2013jha,Arkani-Hamed:2013kca,Arkani-Hamed:2017vfh}, the Cachazo-He-Yuan (CHY) formulation ~\cite{Cachazo:2013hca,Cachazo:2013gna,Cachazo:2013iea} and (ambitwistor) string models ~\cite{Berkovits:2013xba,Mason:2013sva} for the scattering of massless particles in general dimensions, as well as the color/kinematics duality and Bern-Carrasco-Johansson (BCJ) double copy~\cite{Bern:2008qj,Bern:2010ue}. The key ingredient is to consider differential forms in the kinematic space of $n$ massless particles, where we replace color factors of cubic tree graphs in color-dressed amplitudes by wedge-products of $ds$'s, due to a basic observation that they satisfy the same algebra~\cite{Arkani-Hamed:2017mur}. The BCJ color/kinematics duality means that kinematic numerators satisfy Jacobi identities just as the color factors, the form is required to be {\it projective}, which provides a geometric origin for the duality.

Any such projective, scattering forms, can be obtained as pushforward of some differential forms in the moduli space ${\cal M}_{0,n}$, which we call worldsheet forms. This is realized by summing over all solutions of scattering equations, and it is equivalent to CHY formula for the corresponding color-dressed amplitudes. Important examples include the forms for Yang-Mills theory and non-linear sigma model, which are uniquely fixed by gauge invariance and Adler's zero, respectively. Partial amplitudes of the theory from color decomposition can be obtained from the pullback of the form to certain subspaces encoding the ordering~\cite{Arkani-Hamed:2017mur}. These forms, however, are not {\it canonical forms}~\cite{Arkani-Hamed:2017tmz} since they are not $d\log$'s with unit residues, but rather with residues depending on polarizations or other data.

The primary example for canonical forms is the {\it planar scattering form}, $\Omega_{\phi^3} (1,2,\cdots,n)$, which is a $d\log$ scattering form from summing over planar cubic trees respecting the ordering; it represents a bi-adjoint $\phi^3$ amplitude with color stripped for one of the two groups. Geometrically, the pullback of $\Omega_{\phi^3}(1,2,\cdots, n)$ to the subspace for this ordering is the canonical form of an associahedron that is beautifully defined in the kinematic space~\cite{Arkani-Hamed:2017mur}. Moreover, as a geometric reformulation of the CHY formula, the form can be obtained as the pushforward of the Parke-Taylor wordsheet form; the latter is the canonical form of the positive part of moduli space, ${\cal M}_{0,n}^+(1,2,\cdots,n)$, and scattering equations naturally provide a one-to-one map from ${\cal M}_{0,n}^+$ (also an associahedron) to the kinematic associahedron.

Now it is very natural to ask, without the input of physical amplitudes or CHY formulas, what can be said about these scattering forms and corresponding worldsheet forms? More specifically, it is highly desirable to have a mechanism that can {\it generate} both forms in a straightforward way. In this note we will show that, starting from a $(n{-}3)$-dimensional {\it subspace} of the kinematic space, one obtains general scattering forms and worldsheet forms without any other input. This has been implicitly stated in~\cite{Arkani-Hamed:2017mur}: both the planar scattering form and the Parke-Taylor form, can be {\it derived} only from the subspace given by conditions $s_{i,j}={\rm const}$ for non-adjacent $i,j$'s with {\it e.g.} $i,j \neq n$. In sec.~\ref{sec2} we generalize the associahedron story in a systematic way: for any $(n{-}3)$-dimensional subspace, one constructs a scattering form by dressing each cubic tree with the pullback of its wedge product to the subspace, and a worldsheet form from the pullback of scattering equations to it. We will prove that the two forms, though constructed independently, have the remarkable properties that the latter pushes forward to the former. 

The idea that forms can be constructed from subspaces is useful for connecting them to amplitudes, as we show in sec.~\ref{sec322}. Just as color-ordered amplitudes are given by pullback to the subspace for an ordering, it is natural to interpret amplitudes as pullback, which is defined by a pair of subspaces. Such amplitudes are exactly those obtained from BCJ double copy, and they are naturally given by CHY formula associated with the two subspaces. This way of thinking allows one to view scattering equations in a novel way. Given any $(n{-}3)$-dimensional subspace, the scattering equations can be rewritten as a manifestly SL(2)-invariant map from ${\cal M}_{0,n}$ to this subspace; this is obtained by exploting the GL$(n{-}3)$ redundancy of the equations, with the Jacobian given by the worldsheet form. For the case of $\Omega_{\phi^3}(1,2,\cdots,n)$, the map was obtained in~\cite{Arkani-Hamed:2017mur} and we will see that now it naturally generalizes to any subspace. In particular, for an infinite family of subspaces with combinatoric interpretation as spanning trees (called Cayley cases)~\cite{Gao:2017dek}, the map has an elegant form which follows from simple graphic rules. As presented in~\cite{Arkani-Hamed:2017mur}, the forms in these cases are canonical forms of so-called Cayley polytopes in kinematic space (generalizations of the associahedron). We will show in sec.~\ref{cayleyley} that they are naturally derived from the subspaces, and present the explicit construction for these Cayley polytopes.

To illustrate the power of our construction, in sec.~\ref{sec5} we go beyond Cayley cases and discover a much larger class of subspaces that also give $d\log$ forms. The most natural generalization is the so-called ``inverse-soft'' construction, which gives a class of $n$-pt subspaces for $d\log$ forms from any $(n{-}1)$-pt subspace. Concerning the worldsheet forms, our construction corresponds to the well-known ``inverse-soft factor": as we will review, this can be used to recursively build MHV (non-planar) leading singularities in ${\cal N}=4$ super-Yang-Mills (SYM). The most general MHV leading singularities were classified in~\cite{Arkani-Hamed:2014bca}, which include but are not restricted to those constructed using inverse-soft factors. These leading singularities correspond to functions/forms in the moduli space that have nice properties such as factorizations~\cite{Cachazo:2017vkf}. Explicitly, we find very simple subspaces whose worldsheet forms give {\it any} leading singularity functions, with a new formula that is very different from the one in~\cite{Arkani-Hamed:2014bca}. It is intriguing that every MHV leading singularity, viewed as a worldsheet form, now has a scattering form and a (combinatoric) polytope in kinematic space associated with it. All these directly follow from the subspaces we constructed, and we conjecture that they are all simple polytopes.

\section{Scattering forms and worldsheet forms from subspaces}\label{sec2}
In large enough spacetime dimensions, the kinematic space of $n$ massless particles, ${\cal K}_n$, can be spanned by all independent $s_{a b}$'s, thus it has dimension $d:=\frac{n(n{-}3)}{2}$. As shown in \cite{Arkani-Hamed:2017mur}, certain $(n{-}3)$-dimensional subspaces of ${\cal K}_n$ play an important role in the study of scattering forms and in particular canonical forms and positive geometries in kinematic space. In this paper, we initiate a systematic construction, which generalizes the results of~\cite{Arkani-Hamed:2017mur}, for {\it any} generic $(n{-}3)$-dimensional subspace $H$. For any point $x\in H$, we construct two types of closely-related differential forms of dimension $(n{-}3)$: one in the kinematic space ${\cal K}_n$ and the other in the moduli space of $n$-punctured Riemann spheres, ${\cal M}_{0,n}$.

\paragraph{Scattering forms} The differential form in ${\cal K}_n$ is a {\it scattering form}: a linear combination of $d\log$'s of propagators of cubic tree Feynman diagrams with $n$ external legs. Let's denote the collection of all $(2n{-}5)!!$ diagrams as $\Gamma$; each $g\in \Gamma$ is specified by $n{-}3$ Mandelstam variables that are mutually compatible, which are denoted as $s^{(g)}_1, s^{(g)}_2, \cdots, s^{(g)}_{n{-}3}$. We define the {\it wedge product} of $ds$'s for $g$ (the overall sign depends on ordering of $d s $'s):
\be
W_g:=d s^{(g)}_1 \wedge d s^{(g)}_2 \wedge \cdots \wedge d s^{(g)}_{n{-}3} \,.
\ee
It is natural to consider the pullback of such wedge products to $x \in H$:
\be\label{ngngng}
(W_g)|_H:=N_g (x)~(d x_1 
\wedge \cdots \wedge d x_{n{-}3})\,, \qquad N_g(x):=\bigg\lvert\frac{\partial s^{(g)}_i}{\partial x_j}\bigg\rvert
\ee
where $N_g(x)$ is the Jacobian of $W_g$ with respect to $x$'s, which depends on the tangent space of $H$ at $x$. It is possible that all $(2n{-}5)!!$ $N_g$'s vanish at $x$, then we say $H$ is degenerate at $x$; we say the subspace $H$ is non-degenerate if it is non-degenerate everywhere. It is natural to define a scattering form for $H$ at $x$, which is non-vanishing for non-degenerate case
\be\label{form}
\Omega^{(n{-}3)}_H:=\sum_{g\in \Gamma}~N_g~\frac{W_g}{\prod_{i=1}^{n{-}3} s^{(g)}_i} =\sum_{g\in \Gamma}~N_g~\bigwedge_{i=1}^{n{-}3} d\log s^{(g)}_i\,.
\ee
We emphasize that $\Omega^{(n{-}3)}_H(x)$ is completely determined by the tangent space of $H$ at $x$.

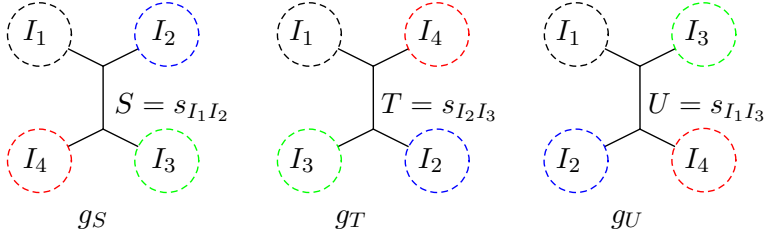
\begin{figure}[!htb]
\begin{center}
\begin{overpic}[width=.8\linewidth]{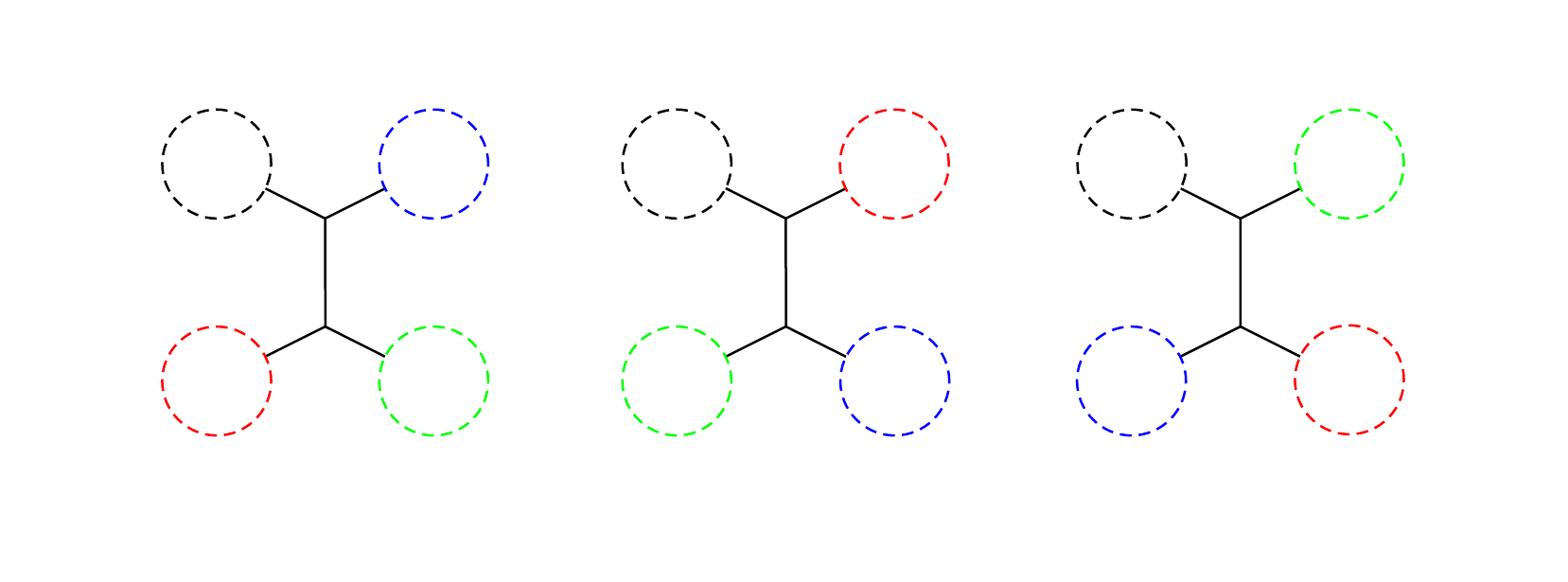}
\put(12,26){$I_1$}\put(26,26){$I_2$}\put(26,12){$I_3$}\put(12,12){$I_4$}
\put(41,26){$I_1$}\put(55,26){$I_4$}\put(55,12){$I_2$}\put(41,12){$I_3$}
\put(70,26){$I_1$}\put(84,26){$I_3$}\put(84,12){$I_4$}\put(70,12){$I_2$}
\put(22,18){$S=s_{I_1I_2}$}\put(51,18){$T=s_{I_2I_3}$}\put(80,18){$U=s_{I_1I_3}$}
\put(18,6){$g_S$}\put(46,6){$g_T$}\put(76,6){$g_U$}
\end{overpic}
\end{center}
\caption{A triplet of three cubic tree graphs that differ by one propagator.}
\label{figjacobi}
\end{figure}

A basic observation in~\cite{Arkani-Hamed:2017mur} is that all linear relations among these wedge products are given by Jacobi identities which are equivalent to those of color factors. There is one such identity for any triplet of graphs, $g_{S}, g_{T}, g_{U}$, that differ only by one propagator, see Figure~\ref{figjacobi}. The wedge products of these three graphs satisfy the Jacobi identity
\be\label{Jacobi}
W_{g_{S}} + W_{g_{T}}+ W_{g_{U}}=\cdots \wedge (d S+ d T + d U)\wedge \cdots=0\,,
\ee
where the distinct Mandelstam variables are $S, T, U$, respectively, and $\cdots$ denote the wedge-products of the remaining $n{-}4$ propagators shared by the three graphs. Denoting the four propagators connecting to the four subgraphs as $s_{I_1}, \cdots, s_{I_4}$, the second equality follows from the basic identity implied by momentum conservation, $d S + d T + d U= d s_{I_1} + d s_{I_2} + d s_{I_3} + d s_{I_4}$. \eqref{Jacobi} implies that \eqref{form} is a {\it projective} form~\cite{Arkani-Hamed:2017mur}, {\it i.e.} it is invariant under a GL$(1)$ transformation $s_I \to \Lambda (s) s_I$ for all subsets $I$ (with $\Lambda(s)$ depending on $s$). As shown in~\cite{Arkani-Hamed:2017mur}, the projectivity of the form is guaranteed if for any three graphs as in Figure~\ref{figjacobi}, we have
\be\label{Jacobi2}
N_{g_S}+ N_{g_T}+ N_{g_U}=0\,.
\ee
Obviously \eqref{Jacobi2} holds, because it is the pullback of \eqref{Jacobi} to $H$ at $x$. Therefore, we have constructed a projective, scattering form given any point in a general subspace.

\paragraph{Worldsheet forms} Similarly, we can study the pullback of scattering equations, $E_a:=\sum_{b\neq a} \frac{s_{a\,b}}{\sigma_a-\sigma_b}=0$ (for $a=1,\cdots, n$),  to $H$, and define its Jacobian with respect to $x_1, \cdots, x_{n{-}3}$:
\be\label{Jac}
J_H(x)=\det{}' \left(\frac{\partial E_a|_H }{\partial x_i}\right):=(r\,s\,t)^{-1} \bigg|\frac{\partial E_a(x, \sigma)}{\partial x_i} \bigg|_{r,s,t}
\ee
where we delete three rows of the derivative matrix, $a=r,s,t$ and compensate with the factor $(r\,s\,t):=\sigma_{r,s}\, \sigma_{s,t}\, \sigma_{t,r}$. It is easy to check that by combining with the top-form of ${\cal M}_{0,n}$, we have a SL$(2)$-invariant $(n{-}3)$-form we call {\it worldsheet form} on ${\cal M}_{0,n}$:
\be
\omega^{(n{-}3)}_H(x)=J_H(x)~d^{n{-}3} \sigma\,, \qquad d^{n{-}3} \sigma:=(i\,j\,k) \prod_{a\neq i,j,k} d \sigma_a
\ee
This is a natural form in ${\cal M}_{0,n}$ associated with the tangent space of $H$ at $x$.

The main claim we make here is
\begin{theo*}
The pushforward of $\omega$ gives $\Omega$:
\be\label{pushforward}
\Omega_H^{(n{-}3)}(x, s)=\sum_{\rm sol.}~\omega^{(n{-}3)}_H(x, \sigma)\,,
\ee
where the sum is over the $(n{-}3)!$ solutions of scattering equations.
\end{theo*}
\noindent  One can show that $J_H(x)$ vanishes if and only if $H$ is degenerate at $x$, in which case \eqref{pushforward} holds trivially. We will prove \eqref{pushforward} in Appendix \ref{appa}, and let's look at the $n=4$ case for now. $H_4$ is one-dimensional and its tangent space can be written as $(d s, d t, d u)=(N_s, N_t, N_u ) dx$ where $N_s+N_t+N_u=0$ guarantees $d s+ d t+ d u=0$. The projective scattering form in ${\cal K}_4$ is given by
\be
\Omega_{H_4}=N_{s}\, d \log \frac{s}{u}+ N_{t}\, d \log \frac{t}{u}
\ee
and the form obtained from the pullback of scattering equations in ${\cal M}_{0,4}$ is
\be
\omega_{H_4}=N_s~d\log \frac{\sigma_{1,2}\sigma_{3,4}}{\sigma_{1,3}\sigma_{2,4}} + N_t~d\log \frac{\sigma_{1,4}\sigma_{2,3}}{\sigma_{1,3}\sigma_{2,4}}\,.
\ee
It is straightforward to see that \eqref{pushforward} holds by plugging in the solution of $n=4$.

In general, $H$ is a hypersurface and both forms are defined locally on it. In the following, we will  consider the special case when $H$ is a {\it hyperplane} that can be defined by $d{-}(n{-}3)$ {\it linear} constraints on the Mandelstam variables. In this case one uses global coordinates $X_1, \cdots, X_{n{-}3}$ for $H$ and any Mandelstam $s$ can be written as a linear combination of $X$'s when pulled back to $H$. For a hyperplane $H$, $N_g$'s become constants (independent of $s$) and the Jacobian becomes a rational function of $\sigma$'s only. Note that different hyperplanes can give identical forms in ${\cal K}_n$ and ${\cal M}_{0,n}$, and we will call them equivalent hyperplanes. In the following we focus on equivalence classes of hyperplanes.

The first and most important example of equivalence classes of $H$ was found in~\cite{Arkani-Hamed:2017mur}. Let $H=h(1,2,\cdots,n)$ denote the class defined by the following constraints: $s_{i\,j}={\rm const}$ for all $d{-}(n{-}3)$ {\it non-adjacent} $i,j$ of a set of $n{-}1$ labels {\it e.g. } $2\leq i, j \leq n$. It has been shown in~\cite{Arkani-Hamed:2017mur} that excluding {\it any} one label from $1, 2, \cdots, n$ gives equivalent hyperplanes, and in view of our construction the results can be summarized as
\begin{itemize}
\item $N_g=\pm 1$ for all planar cubic trees with canonical ordering (with a sign flip for any two trees differ by one propagator),  and $N_g=0$ for non-planar cubic trees.
\item  $J_H={\rm PT}(1,2,\cdots, n)$, thus $\omega$ gives the canonical form of ${\cal M}^{+}_{0,n}(1,2,\cdots,n)$.
\end{itemize}
In this special case, \eqref{pushforward} pushes the Parke-Taylor form, $\omega(1,2,\cdots,n):=\omega_{h(1,2,\cdots,n)}$ to the planar scattering form $\Omega(1,2,\cdots,n):=\Omega_{h(1,2,\cdots,n)}$.

Our construction here can be viewed as a generalization to general (not necessarily $d\log$) forms, both in ${\cal K}_n$ and ${\cal M}_{0,n}$. It is remarkable that they are completely determined by the choice of the hyperplane $H$, without any other inputs. Generally, the meaning of such generalized scattering forms was discussed in~\cite{Arkani-Hamed:2017mur}: they are the dual of {\it color-dressed} amplitudes in certain theories, where $W_g$'s are dual to color factors (the dual of Jacobi identities are given by \eqref{Jacobi}), and $N_g$'s the so-called Bern-Carassco-Johansson (BCJ) numerators that also satisfy Jacobi identities. It is an important open question how to find hyperplanes (or hypersurfaces) $H$ such that the $N_g$'s become BCJ numerators of a given theory, such as Yang-Mills theory (YM) or non-linear sigma model (NLSM)~\cite{Du:2016tbc,Carrasco:2016ldy,Du:2017kpo,Du:2018khm}; equivalently, one can try to find $H$ such that, on the support of scattering equations, the Jacobian $J_H$ equals the reduced Pfaffian, ${\rm Pf}' \Psi_n (\epsilon, k) $ for YM, or $\det' A_n (s)$ for NLSM \cite{Cachazo:2014xea}.

\section{Amplitudes from pullback: BCJ and CHY formulas}\label{sec322}

In the case of planar scattering forms, $\Omega(1,2,\cdots, n)$ represents color-dressed amplitudes for one of the color groups in $U(N) \times U(N')$ bi-adjoint scalar theory, and it is decomposed to the canonical ordering for the other group. Furthermore, the pullback of $\Omega(\alpha)$ to $h(\beta)$ gives the double-partial amplitude $m(\alpha|\beta)$, where $\alpha, \beta$ are the orderings for the two groups. In general, one can study pullback of $\Omega^{(n{-}3)}_H$ to {\it any} hyperplane $H'$. Note that $W_g|_{H'}=N_g' \, (d X'_1 \wedge \cdots \wedge d X'_{n{-}3})$, the pullback reads
\be\label{pullback}
\Omega_H^{(n{-}3)}|_{H'}
=\left(\sum_{g\in \Gamma} \frac{N_g~N'_g}{\prod_{i=1}^{n{-}3} s^{(g)}_i}\right) d X'_1 \wedge \cdots \wedge d X'_{n{-}3} \,,
\ee
and we will call the expression inside the bracket the ``amplitude'' $M_n (H|H')$. This is reminiscent of the BCJ double-copy construction: given two color-dressed amplitudes/forms defined by $H$ and $H'$, the numerators $N_g$ and $N'_g$ satisfy Jacobi identities, and the amplitude for the double-copy $H \otimes H'$ is exactly given by $M_n (H|H')$! ~ Note that for any hyperplane, $M_n(H|H')$ defined from $\Omega_H|_{H'}$ is equal to $M_n (H'|H)$ from $\Omega_{H'}|_{H}$, thus the amplitude is symmetric in $H$ and $H'$. For the special case of $H'=h(\alpha)$ for some ordering $\alpha$, the pullback is equivalent to trace-decomposition of {\it e.g.} $U(N)$ group, as studied in~\cite{Arkani-Hamed:2017mur}. It gives the partial amplitude $M_n(H|h(\alpha))=M_n^{(H)} (\alpha)$, for the form/color-dressed amplitude defined by the subspace $H$. In particular, for $(H,H')=(h(\alpha), h(\beta))$, we recover $m(\alpha|\beta)$.

Moreover, for both $H$ and $H'$, one can define the Jacobian of scattering equations, $J_H(\sigma)$ and $J_{H'}(\sigma)$. An interesting corollary of \eqref{pushforward} is that the double-copy amplitude is given by the CHY formula with $J_H$ and $J_{H'}$:
\be\label{CHY}
M_n (H|H')=\int d^{n{-}3} \sigma~\sideset{}{^{\prime}}\prod_a \delta(E_a)~J_H(\sigma)~J_{H'}(\sigma)\,,
\ee
To prove this, we take the pullback of \eqref{pushforward} to $H'$ and plug in \eqref{pullback} for the RHS:
\be\label{eq33}
\sum_{\rm sol.} d^{n{-}3} \sigma\, J_H\,|_{H'}=\Omega_H|_{H'}=M_n(H | H')~d^{n{-}3} X'\,.
\ee
Now on the LHS we also need to factor out $d^{n{-}3} X'$, which means we want to rewrite the pullback as an integral with delta functions imposing scattering equations:
\be
\sum_{\rm sol.} d^{n{-}3} \sigma\, J_H\,|_{H'}=d^{n{-}3} X'~\int d^{n{-}3}\sigma\,J_H\,\prod_{i=1}^{n{-}3} \delta (X'_i -f'_i (\sigma, s)) \,,
\ee
where the scattering equations inside the delta functions are written as a map from $\sigma$'s to $X'$'s, $X'_i-f'_i(\sigma, s)=0$. Note that, according to \eqref{Jac}, the Jacobian of the transformation from these equations to the standard scattering equations $E_a=0$, is exactly $J_{H'}$,
\be
\prod_i \delta (X_i -f'_i (\sigma, s)=J_{H'}(\sigma) \sideset{}{^{\prime}}\prod_a \delta(E_a)\,,
\ee
from which \eqref{CHY} follows directly! We have seen that $J_H$ arises from the pullback of scattering equations to any hyperplane $H$. Equivalently, it is the Jacobian of fixing a {\it GL$(n{-}3)$ symmetry} of the equations, and the latter can be viewed as a map from ${\cal M}_{0,n}$ to $H$.

Obviously the equations $E_a=0$ are invariant under any $GL(n{-}3)$ transformation: $\sum_a  \Lambda_{i,a} (\sigma) E_a=0$ where $\Lambda_{i,a} (\sigma)$ is a $\sigma$-dependent matrix. We can exploit the symmetry when considering the pullback to $H$; denote the constraints defining $H$ as $L_\alpha(s)=-c_\alpha$ for $\alpha=1,2,\cdots, d{-}(n{-}3)$, and it is obvious that the $d{-}n{+}3$ $L$'s and the $n{-}3$ $X$'s form a basis of ${\cal K}_n$. In this basis, the scattering equations can be written as a $(n{-}3)\times d$ matrix ${\bf C}_{a, (i \alpha)}$ acting on this basis (here $a=1, 2, \cdots n{-}3$ after deleting three equations),
\be
{\bf C}(\sigma) \cdot (X_1, \cdots, X_{n{-}3}, L_1, \cdots, L_{d{-}(n{-}3)})^{\mathrm{T}}=0\,,
\ee
It is clear that if $H$ is non-degenerate, the matrix formed by the first $n{-}3$ columns of ${\bf C}$, denoted as ${\bf \Lambda}$, must be invertible. We can choose its inverse, ${\bf \Lambda}^{-1}$, as a $GL(n-3)$ transformation, and makes the first $n{-}3$ columns the identity matrix by acting ${\bf \Lambda}^{-1}$ on ${\bf C}$. After this transformation, we have ${\bf \Lambda}^{-1} \cdot {\bf C}=(I, U(\sigma))$ where $U_{i, \alpha}=\sum_a \Lambda^{-1}_{i, a} C_{a, \alpha}$ denotes the remaining part for $i=1,2, \cdots n{-}3$ and $\alpha=1,2, \cdots, d{-}(n{-}3)$. We arrive at the {\it scattering-equation map} from ${\cal M}_{0,n}$ to $H$ (recall that  $L_{\alpha}(s)=-c_{\alpha}$)
\be\label{map}
X_i - \sum_{\alpha=1}^{d{-}n{+}3} c_{\alpha}\,U_{i, \alpha} (\sigma)=0\,,
\ee
for $i=1,\cdots, n{-}3$. Since $c$'s are constants, each $X_i$ is expressed as a function of $\sigma$'s, $f_i=\sum_\alpha c_{\alpha} U_{i, \alpha}(\sigma)$, by \eqref{map}. Note that the Jacobian of the transformation depends on those three equations that are deleted, {\it e.g.} $r,s,t$. To obtain a permutation-invariant Jacobian, we can define the $(n{-}3) \times n$ matrix before deletion, ${\bf \Lambda}'$, and the reduced determinant is exactly that given by \eqref{Jac}: $J_H=(r\,s\,t)^{-1} \det {\bf \Lambda'}$. An important point is that the rewriting, \eqref{map}, makes the SL(2)-invariance of scattering equations manifest: each $U_{i, \alpha}$ must be individually invariant under the SL(2) transformation of $\sigma$'s since $X$'s and $c$'s are independent of $\sigma$'s, thus it can only depends on cross-ratios of $\sigma$'s. For a general $H$, these $U$'s can be rather complicated. In the next section, however, we will encounter a class of hyperplanes where~\eqref{map} takes an elegant form with $U$'s given explicitly.

\section{Cayley cases: the rewriting, forms and polytopes}\label{cayleyley}

As proposed in~\cite{Gao:2017dek} and studied in~\cite{Arkani-Hamed:2017mur}, there is a very special class of hyperplanes on which the form $\Omega^{(n{-}3)}_H$, if not zero, can be interpreted as the {\it canonical form} of a convex polytope in ${\cal K}_n$, just like the associahedron for the planar case. These are the so-called Cayley cases, as each of them can be represented by a Cayley tree, or spanning tree of $n{-}1$ labelled vertices. We will see how the Cayley cases naturally arise from the simplest way of rewriting scattering equations as a map, how both forms can be naturally extracted from the tree for which~\eqref{pushforward} can be easily verified, and how to construct polytopes for these cases.

Recall that the kinematic information of the original scattering equations is encoded by $s_{i, j}$'s. By using momentum conservation, one can eliminate {\it e.g.} all $s_{1,i}$'s and write the equations in terms of $\frac{(n{-}1)(n{-}2)}{2}$ $s_{i, j}$'s with $2\leq i<j\leq n$. It is easy to see that the only remaining constraint is that the sum of them vanishes, and a basis of ${\cal K}_n$
can be chosen as any $d$ of them (by eliminating any one of them). To rewrite the equations as a map, the most natural and the simplest way is to choose $H$ such that the $X$'s and $L$'s exactly form such a basis. By choosing $\frac{(n{-}2)(n{-}3)}{2}=d{-}n{+}3$ of $s_{i,j}$'s to be $L$'s (the constants), one can associate the hyperplane $H$ with a {\it graph}, with $(n-2)$ edges $(i,j)$ corresponding to the $n{-}2$ $s_{i,j}$'s as the complement of $L$'s (the variables). Let's denote the graph as $C_n$ and the hyperplane under consideration as $H(C_n)$, and it is clear that any $(n-3)$ of the complement can serve as $X$'s, that is the coordinates of $H(C_n)$. The first claim we make here is that $H(C_n)$ is non-degenerate if and only if $C_n$ is a connected graph! Note that $C_n$ has $n{-}1$ vertices and $n{-}2$ edges, thus $C_n$ must be a tree graph as long as it is connected.

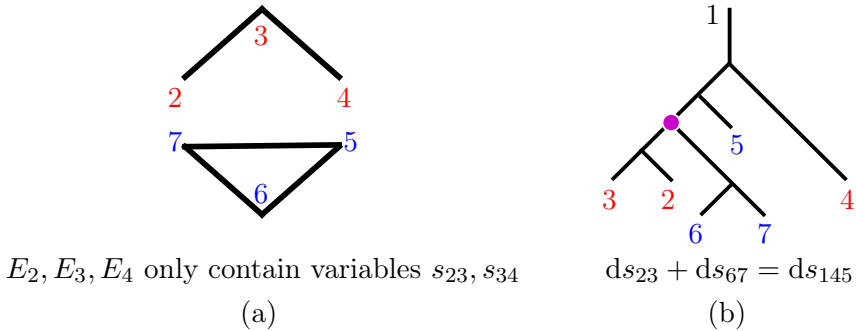
\begin{figure}
\centering
\begin{overpic}[width=0.6\linewidth]
{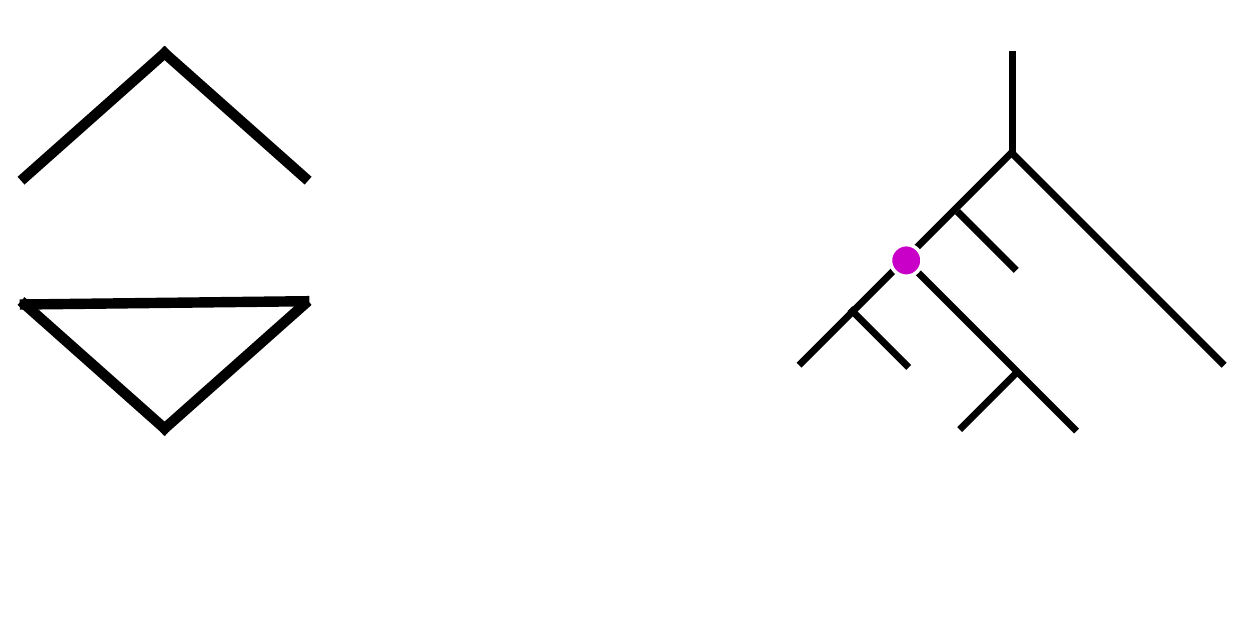}
\put(-0.5,31.5){\textcolor{red}{2}}
\put(12,40.5){\textcolor{red}{3}}
\put(24,31.5){\textcolor{red}{4}}
\put(-0.5,25){\textcolor{blue}{7}}
\put(12,17.4){\textcolor{blue}{6}}
\put(25,25){\textcolor{blue}{5}}
\put(77.5,43.5){1}
\put(10,0){(a)}
\put(77.8,0){(b)}
\put(62.5,16.5){\textcolor{red}{3}}
\put(71,16.5){\textcolor{red}{2}}
\put(75,11.5){\textcolor{blue}{6}}
\put(85,11.5){\textcolor{blue}{7}}
\put(81,24.5){\textcolor{blue}{5}}
\put(97,16.5){\textcolor{red}{4}}
\put(63,6.5){$\mathrm{d}s_{23}+\mathrm{d}s_{67}=\mathrm{d}s_{145}$}
\put(-24,6.5){$E_2,E_3,E_4$ only contain variables $s_{23},s_{34}$}
\end{overpic}
\caption{Examples of vanishing Jacobian and cubic tree.}\label{vnshntree}
\end{figure}

If $C_n$ is disconnected, we will see that both the matrix ${\bf \Lambda}$ and the hyperplane $H(C_n)$ are degenerate. Recall that now $C_n$ must have a connected component which is a tree with no more than $n{-}3$ vertices $i_1,i_2,...,i_m$ (see Figure~\ref{vnshntree}~(a) for example), then $E_{i_1},E_{i_2},...,E_{i_m}$ contain only $m{-}1$ non-constant $s_{i, j}$'s after using momentum conservation to eliminate $s_{1,i}$. This means that there are less than $n{-}3$ independent equations with respect to the $X$'s among the scattering equations, thus the Jacobian $J_{H}=0$, or equivalently the matrix ${\bf \Lambda}$ is not invertible. It is also straightforward to observe that $N_g$ vanishes on $H(C_n)$ for every $g$. Since $C_n$ is disconnected, let's denote the two sets of vertices $A\cup B=\{2,...,n\}$ with no edge between $A$ and $B$. The key is, for every cubic graph $g$, there is at least one vertex that is attached to three edges corresponding to  $s_{T_a},s_{T_b},s_{T_a\cup T_b}$, where $T_a\subset A$ and $T_b\subset B$ (see Figure~\ref{vnshntree}~(b) for example). By pullback to $H(C_n)$ we have
\begin{align}
\label{vnsvnsvns}
d s_{T_{a}}+d s_{T_{b}}=d s_{T_{a}\cup T_{b}}
\end{align}
thus $W_g|_{H(C_n)}=0$. This can also be derived indirectly by using ~\eqref{pushforward} and hence $\omega^{(n-3)}_{H}=0$.

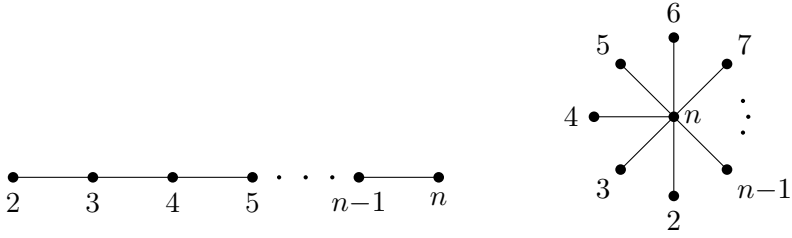
\begin{figure}[!htb]
    \centering
    \subfloat{
   \begin{tikzpicture}[scale=.7]
    \draw[particle] (0,-1) node [below=2pt] {2}--(1.5,-1) node [below=2pt] {3};
   \draw[particle](1.5,-1) --(3,-1) node [below=2pt] {4};
   \draw[particle](3,-1)--(4.5,-1) node [below=2pt] {5};
  \draw[particle] (6.5,-1) node [below=2pt] {$n{-}1$}--(8,-1) node [below=2pt] {$n$};
    \fill (0,-1) circle (.1);
      \fill (1.5,-1) circle (.1);
        \fill  (4.5,-1) circle (.1);
          \fill (3,-1) circle (.1);
         \fill  (6.5,-1) circle (.1);
           \fill  (8,-1) circle (.1);
            \fill  (5,-1) circle (.04);
             \fill  (5.5,-1) circle (.04);
              \fill  (6,-1) circle (.04);
              \node at (3,-2) {~};
    \end{tikzpicture}
         } \subfloat{
   \tikz{\node {~~~~~};}
           }             \subfloat{
   \begin{tikzpicture}[scale=.7]
   \draw[particle] (0,0)node [right=0pt] {$n$}--(-1.5,0) node [left=2pt] {$4$};
   \draw[particle] (0,0)-- (0,1.5) node [above=2pt] {$6$};
 \draw[particle] (0,0)--(0,-1.5) node [below=2pt] {$2$};
   \draw[particle] (0,0)-- (1,1) node [above right=0pt] {$7$};
\draw[particle] (0,0)--(-1,-1) node [below left=0pt] {$3$};
  \draw[particle] (0,0)-- (1,-1) node [below right=0pt] {$n{-}1$};
\draw[particle] (0,0)--(-1,1) node [above left=0pt] {$5$};
      \fill (1.3,0.3)  circle (.04);
          \fill (1.4,0) circle (.04);
          \fill (1.3,-0.3) circle (.04);
           \fill (-1.5,0)  circle (.1);
          \fill (0,0) circle (.1);
          \fill (0,1.5) circle (.1);
           \fill (0,-1.5)  circle (.1);
          \fill (1,1) circle (.1);
          \fill (-1,-1) circle (.1);
           \fill (-1,1) circle (.1);
          \fill (1,-1) circle (.1);
       \end{tikzpicture}
}
\caption{\label{S graph2} Two examples of Cayley graphs, $C^{\rm H}$ and $C^{\rm S}_n$}
\end{figure}

When $C_n$ is connected, it must be a spanning tree of $n{-}1$ vertices, $2,3,\cdots, n$. $H(C_n)$ is defined by $d{-}n{+}3$ constant conditions $s_{p,q}=-c_{p,q}$ where $(p,q)$ is not an edge of  $C_n$; one can choose any $n{-}3$ of the $n{-}2$ $s_{i,j}$'s where $(i,j)$ is an edge as the coordinates of $H(C_n)$. For example, in \cite{Gao:2017dek, Arkani-Hamed:2017mur} two extreme Cayley cases are the linear (or Hamiltonian) tree $C^{\rm H}_n$ and the symmetric, star-shaped tree $C^{\rm S}_n$, which are illustrated in Figure~\ref{S graph2}.

\begin{figure}[!htb]
\centering
\begin{overpic}[width=0.6\linewidth]
{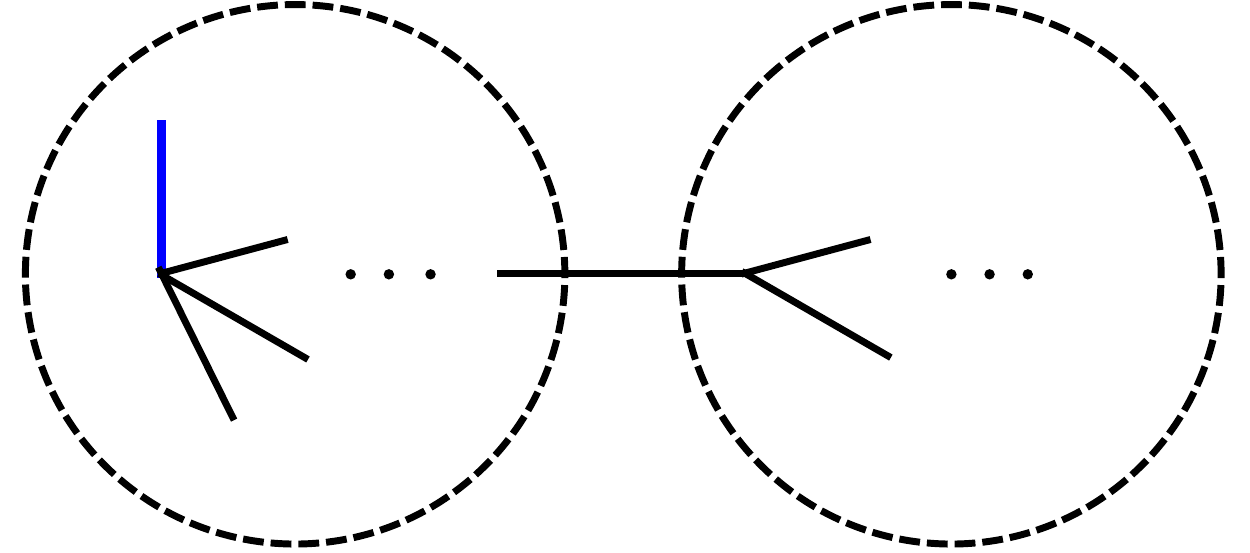}
\put(9.5,32){$r$}
\put(8.5,21.5){$w$}
\put(38,24){$i$}
\put(58,24){$j$}
\put(21.5,4){$L$}
\put(74,4){$R$}
\end{overpic}
\caption{Illustration for a Cayley graph and the rewriting of scattering equations}\label{clynot}
\end{figure}
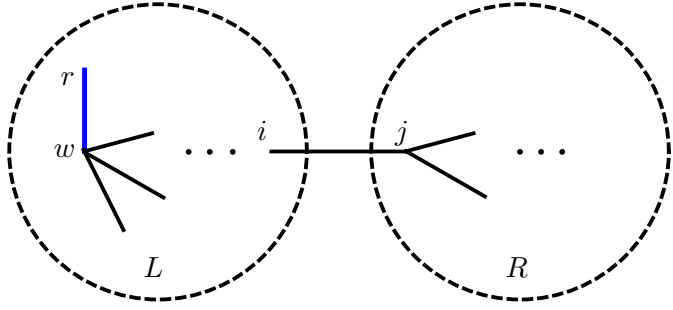

\paragraph{Scattering-equation map and Cayley functions} Now we are ready to derive an elegant rewriting of scattering equations as a map, \eqref{map}, as well as the Jacobian, $J_{H(C_n)}$. As already studied in \cite{Gao:2017dek}, for the Cayley case without label $1$, it is convenient to work in the SL(2) fixing $\sigma_1 \to \infty$. We will rewrite scattering equations, one for each of the $n{-}2$ edges of $C_n$; each edge, $(i,j)$, divides $C_n$ into two parts $L_{(i,j)}$ and $R_{(i,j)}$ (we will omit the subscript $(i,j)$ and our convention is that $i\in L$, $j\in R$), see Figure~\ref{clynot}. Let's take the sum of scattering equations $E_a$ with $a\in L$:
\be
{\cal E}_{(i,j)}:=\sum_{a\in L} E_a=\sum_{\substack{a\in L\\b \neq a, b=1}}^n \frac{s_{a, b}}{\sigma_{a, b}}\,.
\ee
It is interesting to see that all terms with both $a,b \in L$ cancel in this sum, and the remaining ones include $(a,b)=(i, j)$ and those $(a,b)=(p, q)$'s with $p\in L$, $q\in R$:
\be
0={\cal E}_{(i,j)}=\frac{s_{i,j}}{\sigma_{i,j}} + \sum_{p \in L, q\in R} \frac{s_{p,q}}{\sigma_{p,q}}\,.
\ee
By multiplying $\sigma_{i,j}$ and plugging in $s_{p,q}=-c_{p,q}$ we have the scattering-equation map: \be\label{mapCayley}
s_{i,j}-\sum_{p \in L, q\in R} \frac{\sigma_{i,j}}{\sigma_{p,q}} c_{p,q}=0
\ee
for all $n{-}2$ edges $(i, j)$ of $C_n$. One can easily recover the SL(2) invariance and the coefficients of $c_{p,q}$'s are nice cross-ratios of $\sigma$'s:
\be
s_{i,j}-\sum_{p \in L,\,q\in R} \frac{\sigma_{i,j} \sigma_{1,p} \sigma_{1, q}}{\sigma_{p,q} \sigma_{1, i} \sigma_{1,j}}~c_{p,q}=0\,.
\ee
where $n{-}3$ $s_{i, j}$'s serve as the $X$ coordinates of $H(C_n)$, thus \eqref{mapCayley} provides a map from ${\cal M}_{0,n}$ to $H(C_n)$.

Similar to the case of scattering equations, only $n{-}3$ of these $n{-}2$ equations are independent, which can be obtained by deleting any edge, say $(r, w)$, of $C_n$ (see Figure~\ref{clynot}). By arranging the $n{-}3$ equations, ${\cal E}_{(i,j)}$, in an appropriate order, the transformation matrix from $\{E_a\}$ to $\{{\cal E}_{(i,j)}\}$ is a unit triangular matrix with unit determinant, thus the computation of Jacobian simplifies to the product of $n{-}3$ factors $1/\sigma_{i,j}$'s:
\be
\label{eqnclfctn}
J_{H(C_n)}=\det{}'\frac{\partial E}{\partial X}=\det{}' \frac{\partial {\cal E}}{\partial s}
=\frac{1}{(1\,r\,w)} \prod_{n{-}3~{\rm edges}~(i,j)}\frac{1}{\sigma_{i,j}}
\ee
where the product is over $n{-}3$ edges $(i, j) \neq (r, w)$. By ignoring the infinity pre-factor $\sigma_{1, r} \sigma_{w, 1}$ in $(1\,r\,w)$, we arrive at the {\it SL(2)-fixed Cayley function} of~\cite{Gao:2017dek}:
\be\label{fixJ}
J^{\rm fixed}_{H(C_n)}=\prod_{(i,j)~{\rm edges~of}~C_n} \frac 1 {\sigma_{i,j}}
\ee
and the SL(2)-invariance can be recovered by dressing with the prefactor $\prod_{a=2}^n \sigma_{1,a}^{v(a){-}2}$~\cite{Gao:2017dek}. It follows that the worldsheet form $\omega_{H(C_n)}:=d^{n{-}3} \sigma~J_{H(C_n)}$ can be nicely written as
\be\label{caydlog}
\omega_{H(C_n)}^{\rm fixed}=\bigwedge_{n{-}3~(i,j)} d\log {\sigma_{i,j}}
 \iff \omega_{H(C_n})=\bigwedge_{n{-}3~(i,j)} d\log \frac{\sigma_{i,j}\sigma_{1,r} \sigma_{1,w}}{\sigma_{r,w} \sigma_{1,i} \sigma_{1,j}}\ee

For example (see Figure~\ref{S graph2}), for $C^{\rm H}_n$ the rewriting \eqref{mapCayley} was obtained in~\cite{Arkani-Hamed:2017mur}:
\be
L=\{2,\cdots, i\}\,,\quad R=\{i{+}1, \cdots, n\}:\qquad s_{i,i{+}1}-\sum_{p\leq i, i{+}1\leq q} \frac{\sigma_{i, i{+}1} \sigma_{1, p} \sigma_{1, q}}{\sigma_{p, q} \sigma_{1, i} \sigma_{1, i{+}1}} c_{p,q}=0\,, 
\ee
and the Cayley function is the Parke-Taylor factor. For $C^{\rm S}_n$ with {\it e.g. } label $n$ at the center, the rewriting and the (fixed and invariant form of) Cayley function read:
\ba
&&L=\{i\}\,,\quad R=\{2, \cdots, n\}/\{i\}:\qquad s_{i, n}-\sum_{\substack{p=i, q\neq i\\ 2\leq q \leq n{-}1}} \frac{\sigma_{i, q}\sigma_{1, n}}{\sigma_{i, n} \sigma_{1, q}} c_{i, q}=0\,, \nonumber\\
&&J^{\rm fixed}_{C^{\rm S}}=\prod_{i=2}^{n{-}2} \frac 1 {\sigma_{i, n}}\,, \qquad J^{\rm inv.}_{C^{\rm S}}=\frac {\sigma_{1, n}^{n{-}4}} {\prod_{i=2}^{n{-}1} \sigma_{1, i} \sigma_{i, n}}\,.
\ea
where in the first equation we have used $p=i$ and the cross-ratios simplify.

\begin{figure}[!htb]
\centering
\begin{overpic}[width=0.6\linewidth]
{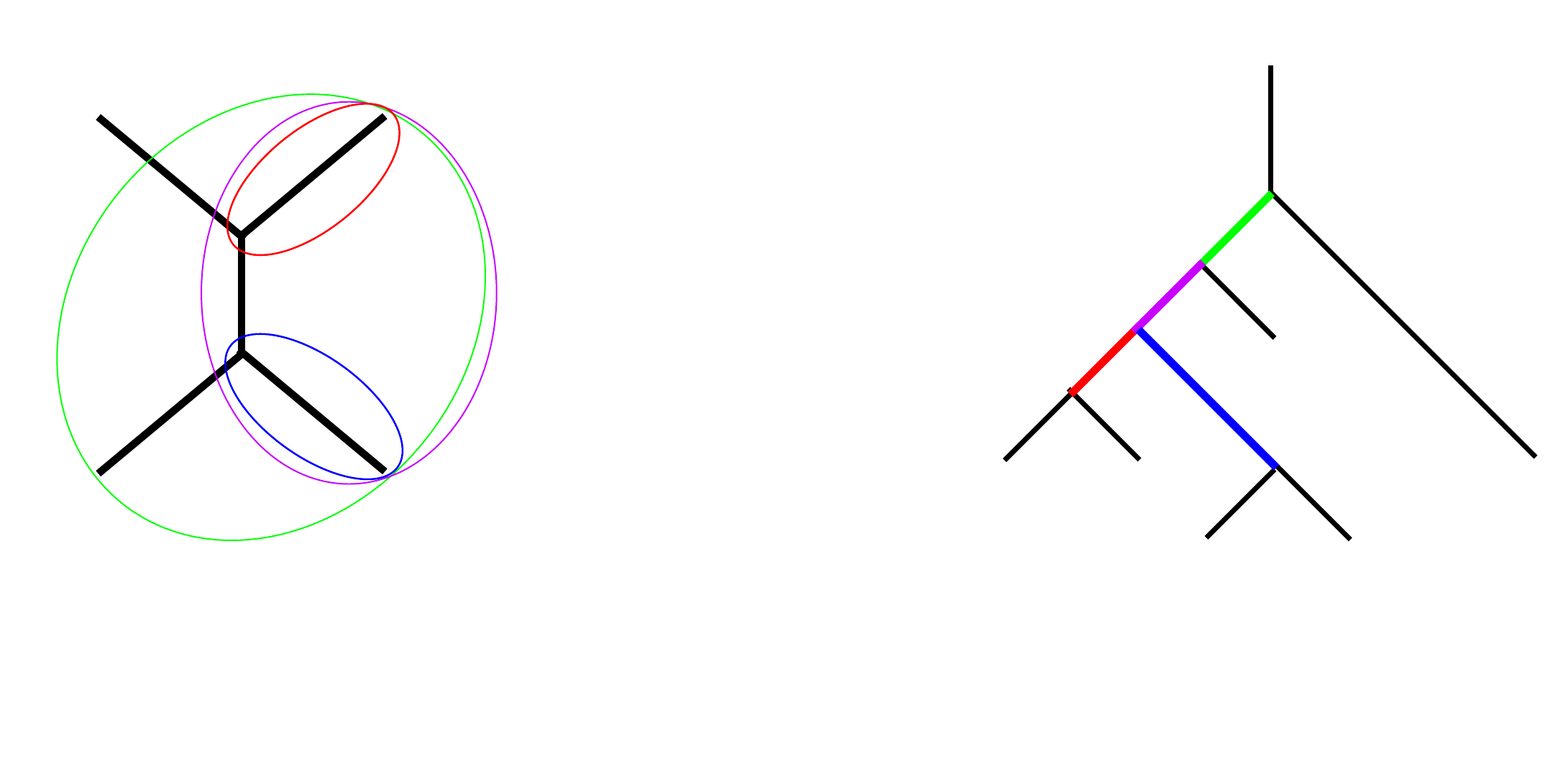}
\put(25,42){2}
\put(3.5,42){4}
\put(12,31.5){3}
\put(25,17){7}
\put(3.5,17){5}
\put(12,27){6}
\put(77.5,43.5){1}
\put(62.5,16.5){3}
\put(71,16.5){2}
\put(75,11.5){6}
\put(85,11.5){7}
\put(81,24.5){5}
\put(97,16.5){4}
\put(-24,6){$W_g\vert_{H}=\mathrm{d}s_{23675}\wedge\mathrm{d}s_{2367}\wedge\mathrm{d}s_{23}\wedge\mathrm{d}s_{67}=(\mathrm{d}s_{2367}+\mathrm{d}s_{56})\wedge\mathrm{d}s_{2367}\wedge\mathrm{d}s_{23}\wedge\mathrm{d}s_{67}$}
\put(-12.6,-1){$=\mathrm{d}s_{56}\wedge\mathrm{d}s_{2367}\wedge\mathrm{d}s_{23}\wedge\mathrm{d}s_{67}=\cdots=\mathrm{d}s_{56}\wedge\mathrm{d}s_{36}\wedge\mathrm{d}s_{23}\wedge\mathrm{d}s_{67}$}
\end{overpic}
\caption{Illustration of non-zero pullback of a cubic tree to $H(C_n)$.}\label{eqpullbc}
\end{figure}

\paragraph{Scattering forms and Cayley polytopes}\label{guaguagua} Now we proceed to scattering forms and the pushforward for Cayley cases, as already studied in~\cite{Arkani-Hamed:2017mur, Gao:2017dek}. It is straightforward to show that the projective, scattering forms for any Cayley graph agree with previous results. If a cubic graph has any pole $s_A$  with subset $A$ that is not a {\it connected subgraph} of $C_n$, one can show that the pullback to $H(C_n)$ gives zero, by a argument similar to \eqref{vnsvnsvns}.  Therefore, any tree that has non-zero pullback consists of $n{-}3$ poles corresponding to compatible, connected subsets of $C_n$ (except for trivial cases $|A|=1$ or $n{-}2$)~\cite{Gao:2017dek}, and let's denote the collection of such cubic trees as $\Gamma_{C_n}$. Furthermore, by choosing any $n{-}3$ $s_{i,j}$ that span $H(C_n)$ as above, the pullback of such $W_g$ always give $\pm 1$, and we have a $d\log$ projective form (see Figure~\ref{eqpullbc} for example):
\ba
&&W_g|_{H(C_n)}=
(\pm) \bigwedge_{(i,j)} d s_{i,j}\,,\quad g\in \Gamma_{C_n}\,,\nonumber\\
&&\Omega_{H(C_n)}=\sum_{g\in \Gamma_{C_n}}~(\pm)~\bigwedge_{{\rm compatible}~A} d\log s_A=\sum_{g \in \Gamma_{C_n}}~(\pm)~\frac{W_g}{\prod_A s_A}\,,
\ea
On the other hand,
as shown in \cite{Gao:2017dek, Cachazo:2017vkf}, the Jacobian $J_{H(C_n)}$ has unit leading singularity at each 0-dimension boundary of $\mathcal{M}_{0,n}$ where every set of pinching punctures belong to a connected subset $A$ of $C_n$. This discussion implies that the pushforward of $\omega_{H(C_n)}^{(n-3)}$ coincides with $\Omega_{H(C_n)}^{(n-3)}$. By construction, the form is projective thus we do not even need to further check the sign of each term in the verification of \eqref{pushforward} for any hyperplane $H(C_n)$.

We have seen that the Cayley case has the advantage that the both the rewriting, \eqref{map} and the proof of~\eqref{pushforward} are simple and clear. Furthermore, there is a nice geometric construction based on $H(C_n)$, which was first discussed in~\cite{Arkani-Hamed:2017mur}. When $C_n$ is a Cayley graph, let's define the top-dimensional, positive region $\Delta$ with all poles being positive:
\begin{align}
\label{rgrgrgrgraaa}
 \Delta(C_n):=\{s_{A}\ge 0,\textrm{ for every nontrivial connected subgraph $A$ of $C_n$}\}
\end{align}
Now by requiring all the constants in defining $H(C_n)$, $c_{p,q}=-s_{p,q}$ being positive, one can check that $\Delta$ have non-empty intersection with $H(C_n)$, which turns out to be a convex polytope that we call {\it Cayley polytope}:
\begin{align}
\label{pcpcpcpcpcaaa}
P(C_n):=\Delta(C_n)\cap H(C_n)\,.
\end{align}
For example, for the star-shaped tree, $C_n^{\rm S}$, we have the region and subspace
\ba
&&\Delta(C_n^{\rm S})=\{s_{i_1 \cdots i_m n } \geq 0\quad | \quad 2\leq i_1, \cdots, i_m\leq n{-}1,~m=1,\cdots, n{-}3\} \,,\nonumber\\
&&H(C_n^{\rm s})=\{ -s_{i j}=c_{i j}>0 \quad | \quad 2\leq i,j \leq n{-}1\}\,,
\ea
and the intersection gives $P(C^{\rm S}_n)$ as the permutohedron polytope~\cite{Arkani-Hamed:2017mur}. We will prove the claim for general Cayley polytope in \ref{appb} by studying the geometric factorization: any codimension-$k$ boundary of $P(C_n)$ corresponds to a set of $k$ compatible poles $s_{A_1},...,s_{A_k}$ where each $A_i$ is a connected subset in $C_n$. Moreover, the canonical form of $P(C_n)$ coincides with the pullback of $\Omega_{H(C_n)}$ on $H(C_n)$, which naturally follows from our construction.

\paragraph{Regions in ${\cal M}_{0,n}$ and relations to graph associahedron}

We have seen that a convex polytope can be constructed beautifully in the subspace of ${\cal K}_n$ for each Cayley case, now we show how to get the same combinatoric polytope as a region in ${\cal M}_{0,n}$. The Cayley worldsheet form, $\omega_{H(C_n)}$ is the canonical form of the region, which can be pushed forward to yield the canonical form of Cayley polytope.

The region can be understood as the union of ${\cal M}_{0,n}^+$ with different orderings in a natural way, following the results of~\cite{Gao:2017dek}. To do this, we need to regard the spanning tree as a directed graph, which also fix the sign convention for $J_{H(C_n)}$. We pick e.g. $n$ as the root and define $C_n$ as a directed graph with all arrows pointing towards $n$. Now the sign convention in $J_{H(C_n)}$ (which we have not been careful about) is that we have $\sigma_j-\sigma_i$ for every edge from $i$ to $j$. Interestingly, we have a nice region that goes with this directed graph:
\be\label{cayre}
R(C_n):=\bigcup_{\substack{\pi \in S_{n{-}2}\\ \pi^{-1} (i)<\pi^{-1}(j)}}{\cal M}_{0,n}^+(1, \pi(2), \cdots, \pi(n{-}1), n)\,.
\ee
It is the union of associahedra with orderings $(1, \pi(2), \cdots, \pi(n{-}1), n)$ such that $i$ precedes $j$ in $\pi$ for each directed edge from $i$ to $j$. For instance, $R(C_{n}^{\mathrm{H}})$ is just the positive part $\mathcal{M}_{0,n}^{+}(1,2,\cdots,n)$ since the directed edges in Hamilton graph are those form $2$ to $3$, $3$ to $4$, and so on. Another example is $C^{\rm S}_n$, all $\pi \in S_{n{-}2}$ contribute in this case since the only directed edges are those from $i$ to $n$ for any $i$, thus $R(C^{\rm S}_n)$ is the union of $(n{-}2)!$ associahedra. It is the following non-trivial identity derived in~\cite{Gao:2017dek} that guarantees that the canonical form of $R(C_n)$ is given by the worldsheet form $\omega_{H(C_n)}$:
\be\label{416eq}
J(C_n)=\sum_{\substack{\pi \in S_{n{-}2}\\ \pi^{-1} (i)<\pi^{-1}(j)} } {\rm PT}(1, \pi(2), \cdots, \pi(n{-}1), n)\,,
\ee
Of course we can choose another label as root which will result in a different region, but they all have the same canonical form (up to a possible sign). In general these regions do not look like a convex polytope in ${\cal M}_{0,n}$, but $R(C_n)$ has exactly the same boundary structure as the corresponding Cayley polytope, $P(C_n)$. For example, the boundaries of $R(C^{\rm S}_n)$ is exactly those of the permutohedron $P(C^{\rm S}_n)$. One can show this by noting that any co-dimension 1 boundary of $R(C_n)$ corresponds to a subset of $\sigma_i$ for $i\in I$ pinching together where $I$ induces a connected subgraph in $C_n$, and so on.

Furthermore, one can show that the scattering-equation map \eqref{mapCayley} maps all boundaries of $R(C_n)$ to corresponding boundaries of $P(C_n)$.  In particular, it is obvious from \eqref{mapCayley} that the boundaries of $R(C_n)$ with $\sigma_i \to \sigma_j$ are mapped to  those of $P(C_n)$ with $s_{i,j}\to 0$. However, unlike the associahedron case, for any $R(C_n)$ that consists of more than one associahedron, its interior is not mapped to the interior of $P(C_n)$ (let alone any one-to-one map). We expect that instead the image of the $R(C_n)$ is the exterior, or the complement of $P(C_n)$ in the subspace $H(C_n)$. This of course explains why the form obtained from pushforward of $\omega_{H(C_n)}$ gives $\Omega_{H(C_n)}$, which is the canonical form for the ``exterior'' as well!

Last but not least, the combinatoric polytopes for $R(C_n)$ are special cases of the so-called graph associahedra~\cite{MichaelCoxeter}, which are natural generalizations of associahedron and play an important role in Coxeter complexes {\it etc.}  To see this, consider a graph $\Gamma(C_n)$ with $n{-}2$ vertices one for each edge $(i,j)$ of $C_n$, and two vertices are connected iff they are adjacent in $C_n$ ({\it i.e.} they share a vertex). For example, $\Gamma(C_n^{\rm H})$ is a Hamilton graph and $\Gamma(C_n^{\rm S})$ a complete graph, with $n{-}2$ vertices. Our $R(C_n)$ and $P(C_n)$ are combinatorially the same polytope as the graph associahedron obtained from $\Gamma(C_n)$. On the other hand, there are of course graphs that cannot be obtained from a spanning trees in this way. For example, we have seen that in rewriting the scattering equations, we encounter disconnected graphs that correspond to degenerate $H(C_n)$. They still give perfectly well-defined $\Gamma(C_n)$ and graph associahedra (for example, the cylcohedron for $n>5$ belong to this case), but there is no Cayley polytope for such cases. Thus our construction singles out a special class of graph associahedra that have a nice realization in kinematic space and scattering-equation maps.

\section{Beyond Cayley: $d\log$ subspaces and leading singularities}\label{sec5}

We have studied Cayley cases in detail, where the entire construction is dictated by a spanning tree 
and the resulting scattering and worldsheet forms are both $d\log$ forms. As already mentioned in~\cite{Arkani-Hamed:2017mur}, such forms are the most direct generalizations of the planar scattering form. We believe the most general $d\log$ scattering and worldsheet forms can be constructed from the so-called {\it $d\log$ subspace} (or hyperplane), as we define now.

A non-degenerate $H$ is a $d\log$ hyperplane if for all $g\in \Gamma$, non-zero $N_g$'s are all equal to each other up to a sign.  According to \eqref{Jacobi2}, the necessary and sufficient condition for a non-degenerate $H$ to be $d\log$ is that for all triplets of graphs as in Figure \ref{figjacobi} where $\{N_{g_s}, N_{g_t}, N_{g_u}\}$ are not all zero, exactly one of the three vanishes (thus the remaining two add up to zero); we further restrict to connected case, thus all non-vanishing ones should be related via such triplets. Given a $d\log$ hyperplane, it is natural to choose its coordinates $X$'s to be $s_i^{(g^{\ast})}$ ($i=1,2,\cdots, n{-}3)$ of a tree $g^{\ast}$ with non-vanishing $N_{g^{\ast}}$, then $N_{g^{\ast}}=1$ and any non-vanishing $N_g=\pm 1$ by definition. Denote the set of trees with non-vanishing $N_g$'s as $\Gamma_H$, then the $d\log$ scattering form for $H$ reads
\be\label{dlogform}
\Omega^{(n{-}3)}_H=\sum_{g \in \Gamma_H} {\rm sgn}(g) \bigwedge_{i=1}^{n{-}3} d\log s_i^{(g)}
\ee
with ${\rm sgn}(g)=N_g=\pm 1$. Similarly the worldsheet form $\omega_H$ is also a $d\log$ form, {\it i.e.} it has unit leading singularities on ${\cal M}_{0,n}$. Instead of fully classifying these $d\log$ hyperplanes, here we focus on a class of $d\log$ $H$ that has a particularly nice interpretation; namely the Jacobians $J_H$ are in one-to-one correspondence with non-planar Maximally-Helicity-Violating (MHV) leading singularities of ${\cal N}=4$ SYM~\cite{Arkani-Hamed:2014bca}.

\paragraph{Review of leading singularities} Recall that a generic top-dimensional MHV on-shell diagram can be characterized by $n{-}2$ triplets of labels $(i_a, j_a, k_a)$ for $1\leq a\leq n{-}2$ (we assume that all labels $1,2,\cdots, n$ are covered). The diagram evaluates to a leading singularity, ${\cal LS}(\lambda_1, \cdots, \lambda_n)$ which is defined on G$(2,n)$ with weight $-2$, {\it i.e.} ${\cal LS} \to \prod_{i=1}^n x_i^{-2} {\cal LS}$ for $\lambda^\alpha_i \to x_i \lambda^\alpha_i$ with $a=1,2, \cdots, n$\footnote{Since this is MHV sector, after stripping off the overall factor $\delta^{4|8} (\sum_{i=1}^n \lambda_i (\tilde\lambda_i |\tilde\eta_i ))$, ${\cal LS}$ only depends on the holomorphic spinors, $\lambda_1, \cdots, \lambda_n$ which form a G$(2,n)$.}. Such functions on G$(2,n)$ are trivially related to the so-called leading singularity functions \cite{Gao:2017dek,Cachazo:2017vkf} on ${\cal M}_{0,n}$ by factoring out an overall factor: ${\bf LS}(\sigma):=\prod_{i=1}^n t_i^2 {\cal LS}(\lambda)$, where $(\lambda_i^{\alpha=1}, \lambda_i^{\alpha=2})=t_i (1,\sigma_i)$ (thus $\langle \lambda_i \lambda_j \rangle=t_i t_j \sigma_{i,j}$). The main result of \cite{Arkani-Hamed:2014bca} is that each ${\bf LS}(\{i,j,k\})$ is given by a nice formula:
\begin{equation}\label{auwofhls}
\mathbf{LS}(\{i,j,k\})= \frac{(\det |\Psi|_{pq}/\s_{pq})^{2}}{\prod_{a=1}^{n-2}(i_aj_ak_a)}
\end{equation}
where $\Psi$ is a $(n-2)\times n$ matrix which only has non-zero entries $\s_{j_{a}k_{a}}$, $\s_{ k_{a}i_{a}}$, and $\s_{ i_{a}j_{a}}$ at the $i_{a}$-th, $j_{a}$-th and  $k_{a}$-th columns of the $a$-th row, respectively.
For example, the matrix $\Psi$ corresponding to the triplet set $\{(1,2,3),\,(3,4,5),\,(5,6,1),\,(2,4,6)\}$ is
\begin{equation}\label{6ptexample}
\Psi = \left(
\begin{array}{cccccc}
\s_{23}  & \s_{31}  & \s_{12}  & 0&0  &0 \\
0  & 0  &  \s_{45} &\sigma_{53} &\sigma_{34} & 0 \\
\s_{56}  & 0  & 0 & 0 & \s_{61}  & \sigma_{15} \\
0 & \sigma_{46} & 0 & \s_{62} & 0 &\s_{24}
\end{array}
\right) \:.
\end{equation}
In the numerator, one computes the determinant after deleting the $p$-th and $q$-th columns with a compensation factor $\frac{1}{\s_{pq}}$, which makes it independent of the choice. In the denominator, the abbreviation $(ijk)$ was defined in \eqref{Jac}.

From now on, the abbreviation LS is always used to denote the leading singularity function on  ${\cal M}_{0,n}$ \eqref{auwofhls}.  Different triplet sets can give identical LS function, if they are related by an equivalence relation known as the ``square move'': if two triplet sets differ by two triplets, $(i,j,k)$, $(i,j,\ell)$ and $(k,\ell,i)$, $(k,\ell,j)$ respectively, the resulting LS functions are equal up to a sign (see \cite{Arkani-Hamed:2014bca}). It is sufficient to consider any representative triplet sets, {\it e.g.} by choosing $T_n=\{(1,2,3), (1,3,4), \cdots, (1, n{-}1, n)\}$ we have ${\bf LS}(T_n)=PT(1,2,\cdots,n)$.

If a triplet set $T_{n}$ of $n$-pt (or any of its equivalent triplet sets) has a particle, say $n$,  only appearing once, say in triplet $(a,b,n)$, then by removing it the remaining $n{-}3$ triplets, denoted as $T_{n-1}$ describe a $(n-1)$-pt LS without $n$, and they are related by
\begin{equation}
\mathbf{LS}(T_{n}) = \frac{\s_{ab}}{\s_{an}\,\s_{ bn}}~\mathbf{LS}(T_{n-1}) \:.
\label{LSres}
\end{equation}
where the prefactor is known as an ``inverse-soft factor'' for inserting $n$ between $a$ and $b$ (a term  introduced for gauge-theory amplitudes, see~\cite{ArkaniHamed:2012nw}), and we call $\mathbf{LS}(T_{n})$ an inverse-soft (IS) reducible LS function since it can be obtained from the lower point one by multiplying with an inverse-soft factor.

If this procedure can be repeated until only we are left with one triplet for $n=3$, we call such a LS function an IS-constructible one. On the contrary, those that do not admit any IS-reduction are called IS-irreducible LS functions. In the following, we will build hyperplanes whose Jacobians exactly correspond to these LS functions, first for IS-constructible ones and then for general cases.

Like the Cayley cases \eqref{cayre}, there are regions in ${\cal M}_{0,n}$ for any LS function. We can define such a region as the union of associahedra  with different ordering,  where for each ordering all the appropirately chosen triplets are ordered correctly:
\be
R({\mathbf {LS}}(\{i,j,k\})):=\bigcup_{\substack{\rho \in S_{n}/Z_n\\ \rho^{-1} (i)<\rho^{-1}(j)<\rho^{-1}(k)}}{\cal M}_{0,n}^+(\rho(1), \rho(2), \cdots, \rho(n{-}1), \rho(n))\,,
\ee 
whose canonical form ${\mathbf {LS}}~d^{n-3}\sigma$ can be pushed forward to yield a canonical form of a polytope described in the following  in a kinematic space.  This equation above is guaranteed by the non-trivial identity similar to \eqref{416eq} derived in \cite{Arkani-Hamed:2014bca}.

\subsection{Inverse-soft construction for $d\log$ subspaces}\label{sec51}
We first introduce a simple, recursive construction for $n$-point $d\log$ hyperplane from any given $(n{-}1)$-point $d\log$ hyperplane. Recall that a $n$-point hyperplane is defined by $\frac{(n{-}3)(n{-}2)}{2}$ constraints in ${\cal K}_n$, and to build it from a $(n{-}1)$-point hyperplane by adding particle $n$ requires $n{-}3$ more constraints. The simplest way to do so is to impose additional $n{-}3$ constraints of the form $s_{in}={\rm const}$ for $n{-}3$ $i$'s chosen from $\{1, 2, \cdots, n{-}1\}$, {\it i.e.} $i\neq a,b$ for some $a,b$. Remarkably, for any given $d\log$ hyperplane $H_{n{-}1}$ (regardless of how it is constructed), the hyperplane arise from {\it inverse-soft} (IS) construction,
\be
H_n^{(ab)}:=H_{n{-}1} \cup \{ s_{in}={\rm const}| 1\leq i\leq n{-}1, i\neq a,b\}\,,\label{IScondition}
\ee
is also a $d\log$ one! The notation $H_n^{(ab)}$ indicates that, both for the scattering form and worldsheet form, in a precise sense as follows, particle $n$ is inserted between $a$ and $b$. The results here are the following two claims:
\begin{stat*}
\noindent(I): All $n$-point trees that have non-zero pullback to $H_n^{(ab)}$ are those with leg $n$ inserted between leg $a$ and $b$, for any $(n-1)$-point tree with non-zero pullback to $H_{n-1}$; with coordinates for $H_n^{(ab)}$ chosen to be those of $H_{n{-}1}$ with one more $s$, the coefficients remains $\pm1$ after inserting $n$. From these pullbacks we can trivially write the $d\log$ form $\Omega_{H_n}$ by \eqref{form}. \\

\noindent(II): The Jacobians/worldsheet forms for $H_n^{(ab)}$ and $H_{n{-}1}$ are related by:
\begin{equation}\label{ISF}
J(H_n^{(ab)})=\frac{\sigma_{ab}}{\sigma_{an}~\sigma_{bn}}~J(H_{n{-}1})
\iff \omega(H_n^{(ab)})=d\log \frac{\sigma_{an}}{\sigma_{bn}} \wedge \omega(H_{n{-}1})\,,
\end{equation}
where the factor $\frac{\sigma_{ab}}{\sigma_{an}~\sigma_{bn}}
$ is exactly the inverse soft-factor for LS functions in \eqref{LSres}.
\end{stat*}

\begin{figure}[!htb]
 \centering
 \subfloat[\label{ISwithoutn}]{
\begin{tikzpicture}[shorten >=0pt,draw=black,scale=.7,
        node distance = .65cm,
        neuron/.style = {circle, minimum size=3pt, inner sep=0pt,  fill=black } ]
     \node[neuron] (1) {};
     \node[ neuron,right of = 1] (2)  {};
   \node[ neuron,above of = 2] (3)  {};
     \node[ neuron,right of = 2,xshift=1.5cm] (4)  {};
    \node[right of = 4] (8)  {};
     \node[ neuron,right of = 8] (9)  {};
    \node[ neuron,above of = 4] (5)  {};
    \node[ neuron,above of = 5] (6)  {};
     \node[ neuron,right of = 5] (7)  {};
      \node at ($(9)+2*(9)-2*(8)$)  {~};
         \node at ($(1)-2*(9)+2*(8)$)  {~};
 \draw (1)--(2)--(3);
    \draw (6)--(4)--(9);
     \draw (5)--(7);
\draw (2)node[below=0pt,xshift=1.05cm]{$s_{I}$}--(4);
\draw[blue,dashed] (0.5,0.5) ellipse (.8 and 1.2);
\node[blue] at (.5,1.1) {$L$};
\draw[blue,dashed] (4.9,1) ellipse (1.6 and 2.1);
\node[blue] at (5.0,2.1) {${R}$};
\end{tikzpicture}
}
 \subfloat[\label{ISwithn}]{
\begin{tikzpicture}[shorten >=0pt,draw=black,scale=0.7,
        node distance = .65cm,
        neuron/.style = {circle, minimum size=3pt, inner sep=0pt,  fill=black } ]
     \node[neuron] (1) {};
     \node[ neuron,right of = 1] (2)  {};
   \node[ neuron,above of = 2] (3)  {};
     \node[ neuron,right of = 2] (10)  {};
     \node[neuron, above of= 10,yshift=0.3cm] (11) {};
     \node[neuron,right of = 10,xshift=0.5cm] (4) {};
    \node[right of = 4] (8)  {};
     \node[neuron, right of = 8] (9)  {};
    \node[ neuron,above of = 4] (5)  {};
    \node[ neuron,above of = 5] (6)  {};
     \node[ neuron,right of = 5] (7)  {};
      \node at ($(9)+2*(9)-2*(8)$)  {~};
         \node at ($(1)-2*(9)+2*(8)$)  {~};
 \draw (1)--(2)--(3);
    \draw (6)--(4)--(9);
     \draw (5)--(7);
     \draw (10)--(11)node[above=0pt]{$n$};
\draw (2)node[below=0pt,xshift=0.45cm]{$s_L$}--(10);
\draw (10)node[below=0pt,xshift=0.55cm]{$s_R$}--(4);
\draw[blue,dashed] (0.5,0.5) ellipse (.8 and 1.2);
\node[blue] at (.5,1.1) {$L$};
\draw[blue,dashed] (4.5,1) ellipse (1.6 and 2.1);
\node[blue] at (4.4,2.1) {${R}$};
\end{tikzpicture}
}
\caption{}
\end{figure}

To prove {\bf{claim}} (I), recall that any $n$-point tree can be viewed as an $(n{-}1)$-point tree with leg $n$ inserted at a particular propagator, which we denote as $s_I$ (see Figure \ref{ISwithoutn}). The propagator divides the whole tree into two parts and we use $L$ and $R$ to denote the particle sets of two sides respectively. In the $n$-point tree, instead of $s_I$, now we have two propagators $s_L$ and $s_R$, as shown in Figure \ref{ISwithn}.

For the wedge product of $d s$'s of the tree to have a non-vanishing pullback to $H_{n}^{(ab)}$, $a$ and $b$ must be on two sides of $n$, otherwise ({\it e.g.} when $a,b$ are both in $L$):
\begin{equation*}
ds_L\wedge ds_R \Big\rvert_{H_{n}^{(ab)}} = ds_L\wedge (ds_L+\sum_{i\in L} d s_{n i})\Big\rvert_{H_{n}^{(ab)}}=\sum_{i\in L} ds_L \wedge d s_{ni} =0
\end{equation*}
where in the second equality we have used constant conditions (\ref{IScondition}), $d s_{n i}=0$. Thus only trees with $n$ inserted between the leg $a$ and $b$ contribute to $\Gamma_{H_{n}^{(ab)}}$. Let's denote the $(n{-}1)$-pt tree as $g'$, and the $n$-pt tree $g$, with the wedge product of $ds$'s as $W_{n{-}1}^{(g')}$ and $W_{n}^{(g)}$, respectively. Then we find the pullback of $W_{n}^{(g)}$ to $H_{n}^{(ab)}$ can be written as
\begin{align*}
W_{n}^{(g)}\Big\rvert_{H_{n}^{(ab)}} &= W^{(g)}_{L}\wedge ds_L \wedge ds_R \wedge W^{(g)}_{R}\Big\rvert_{H_{n}^{(ab)}} \\
&=\pm  W^{(g)}_{L}\wedge ds_L \wedge W^{(g)}_{R}\Big\rvert_{H_{n}^{(ab)}} \wedge ds_{an}
\end{align*}
where $W_{L}^{(g)}$ and $W_{R}^{(g)}$ are the wedge product of $ds$'s for the left and right part, respectively, and with $d s_{n i}=0$ we have replaced $d s_R$ by $d s_{a n}=- d s_{b n}$ and pull it out to the rightmost (with a possible overall sign). Note that the $ds$'s in $W_{L}^{(g)}$ and $W_{R}^{(g)}$ are independent of $n$ since the $n$-dependent part vanishes in the wedge product. We conclude
\begin{align*}
 W^{(g)}_{L}\wedge ds_{I} \wedge W^{(g)}_{R}\Big\rvert_{H_{n}^{(ab)}} = W^{(g^{\prime})}_{n-1}\Big\rvert_{H_{n-1}} \:,
\end{align*}
where the RHS is in $(n{-}1)$-pt kinematic space, and this completes the proof of {\bf{claim}} (I).

To prove {\bf{claim}} (II), just note that the last scattering equation can be rewritten as%
\begin{equation}
E_{n}=s_{a,n}\left( \frac{1}{\sigma_{n,a}}-\frac{1}{\sigma_{n,b}}\right)
+\sum_{i\neq a,b,n} s_{n,i}\left(\frac{1}{\sigma_{n,i}}-\frac{1}{\sigma_{n,b}}\right) \label{ScaE}
\end{equation}
where $s_{b,n}=-s_{a,n}-\sum_{i\neq a,b,n}s_{i,n}$ has been used. It is easy to see that the second term has no contribution to the Jacobian determinant $J(H_{n}^{(ab)})$ since $s_{n i}$'s are constant. Thus the Jacobian matrix has the form
\begin{equation*}
 \left(\frac{\partial E}{\partial s}\right)_{n}^{\prime} =
 \left( \begin{array}{cccc|c}
 &  &  &  & \ast\\
    \multicolumn{4}{c|}{\left(\frac{\partial E}{\partial s}\right)_{n-1}^{\prime}} & \vdots\\
&  &  &  & \ast \\
\cline{1-4}
0 & 0 & \cdots & \multicolumn{1}{c}{0} & \frac{\sigma_{ab}}{\sigma_{an}\sigma_{bn}}
 \end{array}\right)\:,
\end{equation*}%
where primes mean that three rows have been removed, and the $(n{-}3,n{-}3)$-minor $\operatorname{det}\left(\frac{\partial E}{\partial s}\right)_{n-1}^{\prime}$ is exactly $J(H_{n-1})$. Hence we have
\begin{equation*}
J(H_n^{(ab)})=\frac{\sigma_{ab}}{\sigma_{an}~\sigma_{bn}}~J(H_{n{-}1})\,.
\end{equation*}
Finally, the worldsheet form can be obtained by using the identity $
\frac{\sigma_{a\,b}}{\sigma_{a\,n}\sigma_{b\,n}}d\sigma_{n} = d\log\frac{\sigma_{a\,n}}{\sigma_{b\,n}}$.

By recursively using the IS construction \eqref{IScondition}, one can build a large class of $d\log $ hyperplanes by inserting new particles. We call such hyperplanes IS-constructible hyperplanes, where the $(n{-}2)(n{-}3)/2$ constraints defining $H_n$ are all of the form $s_{i\,j}={\rm const}$. Up to relabeling, we can always assume that the particles being added are $p=4,5,\cdots, n$, and every step we have $H_p^{(a_p, b_p)}$ (with $1\leq a_p, b_p <p$) by adding $p{-}3$ constraints, $s_{i, p}={\rm const}$, for $i \neq a_p, b_p$. Thus any IS-constructible hyperplane is labeled by a sequence of $n{-}3$ pairs $(a_p, b_p)$, and we denote it as $H_n(\{a,b\})$. By (I) we can write down the $d\log$ scattering form $\Omega_{H(\{a,b\})}$ by finding $\Gamma_{H(\{a,b\})}$ and the signs of pullback. By \eqref{ISF} of {\bf{claim}} (II) we have
\be \label{JacofIS}
J_{H(\{a,b\})}=\frac 1{(123)}\,\prod_{p=4}^n \frac{\sigma_{a_p, b_p}}{\sigma_{a_p, p}\,\sigma_{b_p, p}} \implies \omega_{H(\{a,b\})}=\bigwedge_{p=4}^n d\log \frac{\sigma_{a_p, p}}{\sigma_{b_p, p}}\,.
\ee

It is now completely obvious that the Jacobian of IS-constructible hyperplanes are exactly those IS-constructible LS functions. More precisely, the LS characterized by $(1,2,3)$ and $(a_{p},b_{p},p)$ for $4\leq p\leq n$ is the same as (\ref{JacofIS}). Cayley hyperplanes are a special class of IS-constructible ones where all $a_{p}$'s (or $b_{p}$'s) are given by a fixed label. We can choose it to be $a_p=1$ for $p=4,\cdots, n$, and the edges of Cayley trees are chosen to be $(2,3), (a_4, 4), \cdots, (a_n, n)$, which represent most general trees up to relabeling.
\begin{figure}[!htb]
\centering
\includegraphics[scale=.65]{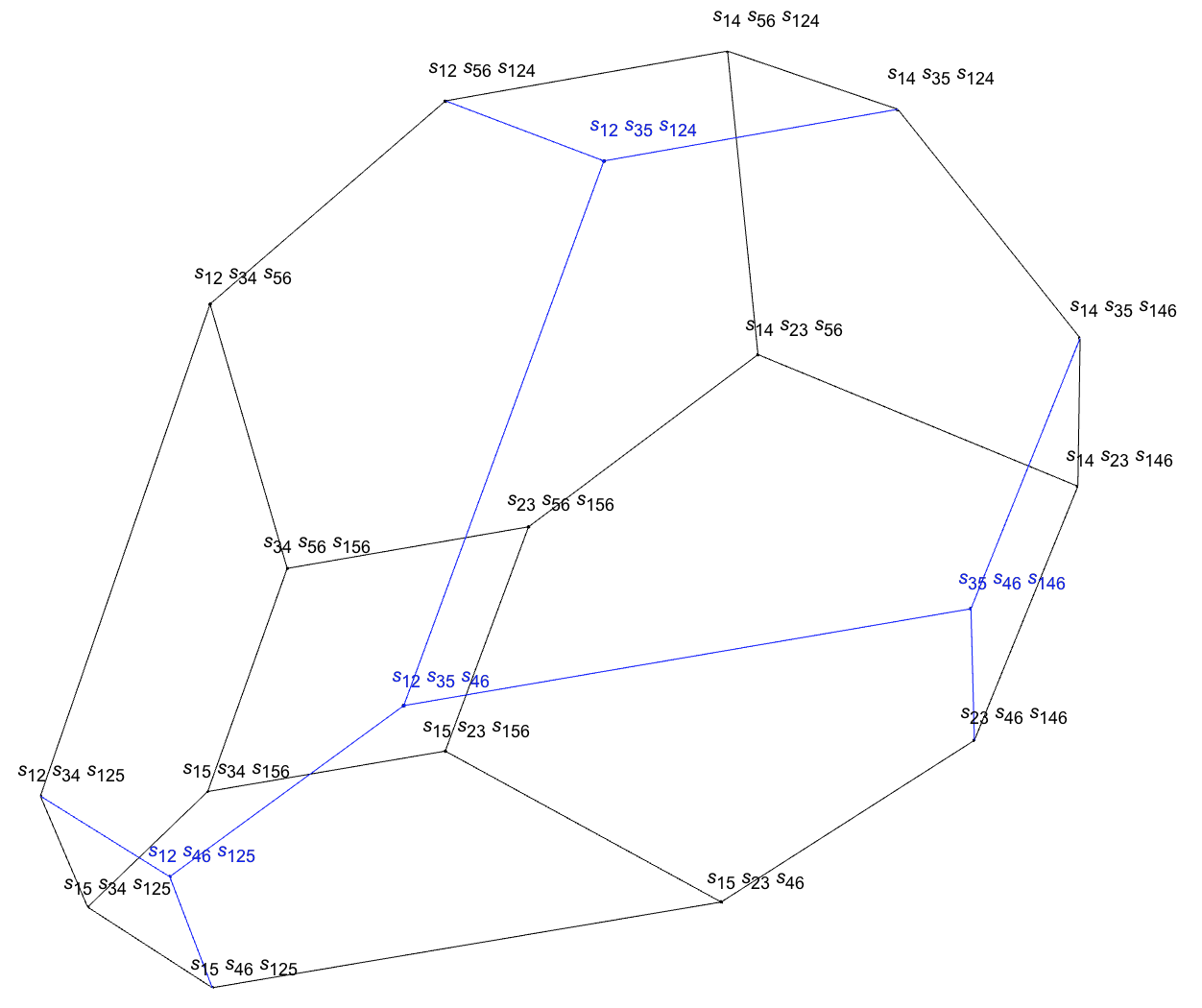}
\caption{Combinatoric polytope  from  $\mathbf{LS}(\{(1,2,3),(1,3,4),(1,3,5),(4,5,6)\})$
\\
Here each vertix corresponds to a trivalent graph charactered by three poles 
\label{22isf}}
\end{figure}
It is trivial to see that up to $n=5$, all LS functions are IS-constructible since the corresponding triplets always have a particle which only appear once. Furthermore, all these LS functions belong to the Cayley cases. For $n=6$, we have the first IS-constructible LS function which is not Cayley function; up to relabelling, its triplets can be chosen to be $\{(1,2,3),(1,3,4),(1,3,5),(4,5,6)\}$, and the constant condition for the corresponding hyperplane is
\begin{equation}
\{s_{24},\, s_{25}, \, s_{45},\,s_{16},\,s_{26},\,s_{36}\text{ are constants}\}.
\end{equation}
The Jacobian for the pullback of scattering equations to this hyperplane is
\begin{equation*}
\frac{\sigma_{13}\sigma_{45}}{\sigma_{12}\sigma_{23}\sigma_{41}
\sigma_{34}\sigma_{15}\sigma_{35}\sigma_{46}\sigma_{56}} \:,
\end{equation*}
Its scattering forms can be computed simliarly, and we obtain the combinatorial polytope corresponding to this LS function, which is not a Cayley polytope (thus differs from those conisdered in \cite{Arkani-Hamed:2017mur} and \cite{Gao:2017dek}), see figure \ref{22isf}.

\subsection{Subspaces for general leading singularity functions}\label{sec52}
Starting $n=6$ we encounter LS functions that are not IS-irreducible, and for $n=6$ up to relabelling there is only one such example, which is also the only one that is not IS-constructible for $n=6$. The triplets have been given in \eqref{6ptexample}, and the LS function reads
\begin{align}
\mathbf{LS}(\{(1,2,3),(3,4,5),(5,6,1),(2,4,6)\}) =
\frac{
(\s_{13}\s_{45}\s_{26}
-\s_{12}\s_{35}\s_{46}
)^2}
{(123)(345)(561)(246)}
\,,\label{guebiw}
\end{align}
where the numerator is not a monomial of $\s_{ij}$ any more. This completes our classification for all types of LS functions for $n=6$. For higher points, there are more and more IS-irreducible LS functions, for example, there are only one and two types of IS-irreducible cases for $n=7$ and $n=8$ respectively, but we find 24 and 205 types of IS-irreducible cases for $n=9$ and $n=10$ respectively (some of their triplets are listed in Table.\ref{tab:irrtripletsets}). The most general LS functions can be obtained from these irreducible ones by applying more IS constructions, \eqref{IScondition}. For example, one can obtain the $n=7$ case $\mathbf{LS}(\{(1,2,3),(3,4,5),(5,6,1)$ $,(2,4,6),\underline{(2,4,7)}\})$ by inserting $7$ between $2,4$, from the $n=6$ irreducible one \eqref{guebiw}. 

Obviously we need a different method for constructing the $d\log$ hyperplane for these irreducible LS functions. A new feature of $d\log$ hyperplanes for irreducible cases is that we will not only have $s_{i,j}={\rm const}$ but also $s_{i,j,k}={\rm const}$. We conjecture that these constant conditions can be read off from the triplets. Before we present our proposition for the most general case, we will first illustrate the result for our $n=6$ example \eqref{guebiw}, and the $n=7$ example as given in Table~\ref{tab:irrtripletsets}.

\begin{table}[t!]
\centering
\begin{tabular}{l|l}
\hline
$n=7$& $\{(1,2,3),(3,4,5),(5,6,7),(7,1,4),(2,4,6)\}\vphantom{\Big(}$\\
\hline
\multirow{2}{*}{$n=8$} &  $\{(7,1,2),(7,3,4),(7,5,6),(8,1,3),(8,2,5),(8,4,6)\}\vphantom{\Big(}$\\
\cline{2-2}
 &  $\{(7,8,1),(7,2,3),(7,4,5),(8,2,4),(8,3,6),(1,5,6)\}\vphantom{\Big(}$ \\
\hline
\multirow{3}{*}{$n=9$} & $\{(1,2,4),(1,5,6),(1,7,8), (2,3,5),(2,8,9), (3,4,6), (3,7,9)\}\vphantom{\Big(}$ \\
\cline{2-2}
& $\{(1,2,3),(1,4,7),(1,5,6),(2,4,9),(2,5,8),(3,6,8),(3,7,9)\}\vphantom{\Big(}$ \\
\cline{2-2}
&  $\quad+ 22\text{ more}\vphantom{\Big(}$   \\
\hline 
\multirow{3}{*}{$n=10$} & $\{(1,2,5),(1,4,6),(1,7,8),(2,3,9),(2,7,10),(3,4,10),(3,5,6),(4,8,9)\}\vphantom{\Big(}$ \\
\cline{2-2}
& $\{(1,2,5),(1,6,7),(1,8,9),(2,3,6),(2,4,9),(3,4,7),(3,8,10),(4,5,10)\}\vphantom{\Big(}$ \\
\cline{2-2}
&  $\quad + 203\text{ more}\vphantom{\Big(}$   \\
\hline
\end{tabular}
\caption{All types of IS-irreducible triplet sets up to relabelling for $n=7$ and $n=8$, and some examples of IS-irreducible triplet sets for $n=9, 10$.}
\label{tab:irrtripletsets}
\end{table}

\begin{table}[t!]
\centering
\begin{tabular}{l|l}
\hline
$n=7$& $(132,264,165,33)\vphantom{\Big(}$\\
\hline
\multirow{2}{*}{$n=8$} &  $(612,1530,1332,468,56)\vphantom{\Big(}$\\
\cline{2-2}
 &  $(624,1560,1360,480,58)\vphantom{\Big(}$ \\
\hline
\multirow{3}{*}{$n=9$} & $(3788,11364,12702,6464,1449,111)\vphantom{\Big(}$ \\
\cline{2-2}
& $(3698,11094,12414,6338,1431,111)\vphantom{\Big(}$ \\
\cline{2-2}
&  $\quad+ 7\text{ more}\vphantom{\Big(}$   \\
\hline 
\multirow{3}{*}{$n=10$} & $(21966,76881,105125,70610,23933,3730,199)\vphantom{\Big(}$ \\
\cline{2-2}
& $(22708,79478,108646,72920,24674,3830,202)\vphantom{\Big(}$ \\
\cline{2-2}
&  $\quad + \text{80 more}\vphantom{\Big(}$   \\
\hline
\end{tabular}
\caption{All $f$-vectors corresponding to triplet sets listed in Table \ref{tab:irrtripletsets}. }
\label{tab:fvector}
\end{table}

\begin{figure}[!htb]
\subfloat[\label{6ptir}for $\mathbf{LS}(\{(1,2,3),(3,4,5),(5,6,1),(2,4,6)\})$]{
\includegraphics[scale=.45]{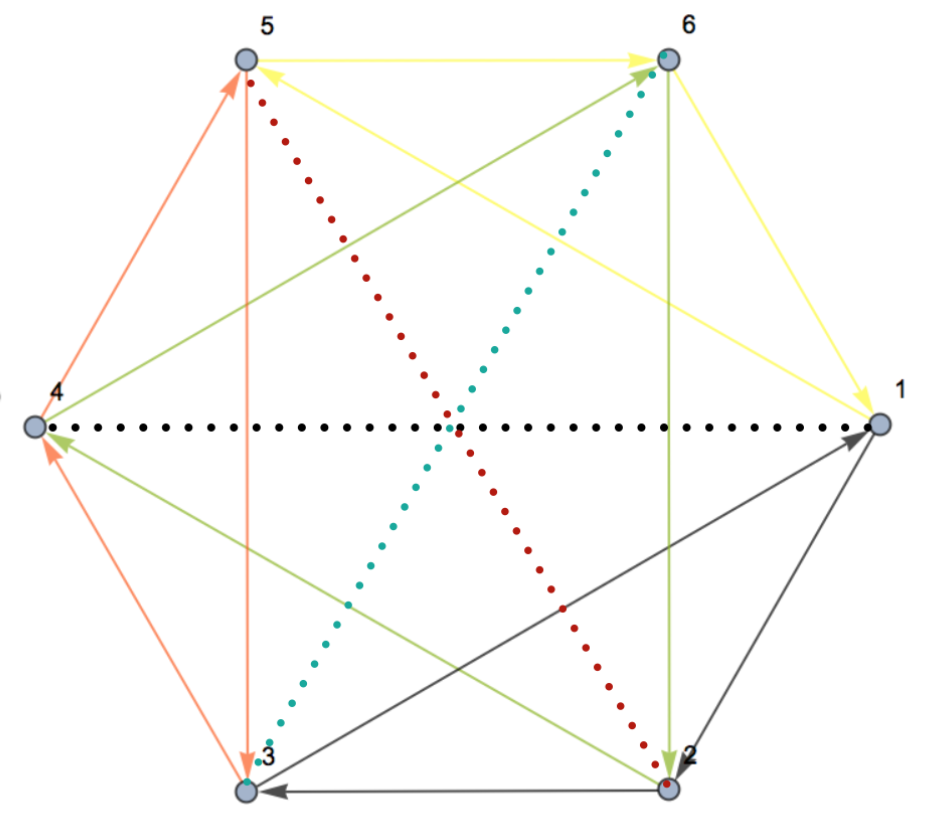}}
\subfloat[\label{7ptir}for $\mathbf{LS}(\{(1,2,3),(3,4,5),(5,6,7),(7,1,4),(2,4,6)\})$]{
\includegraphics[scale=.34]{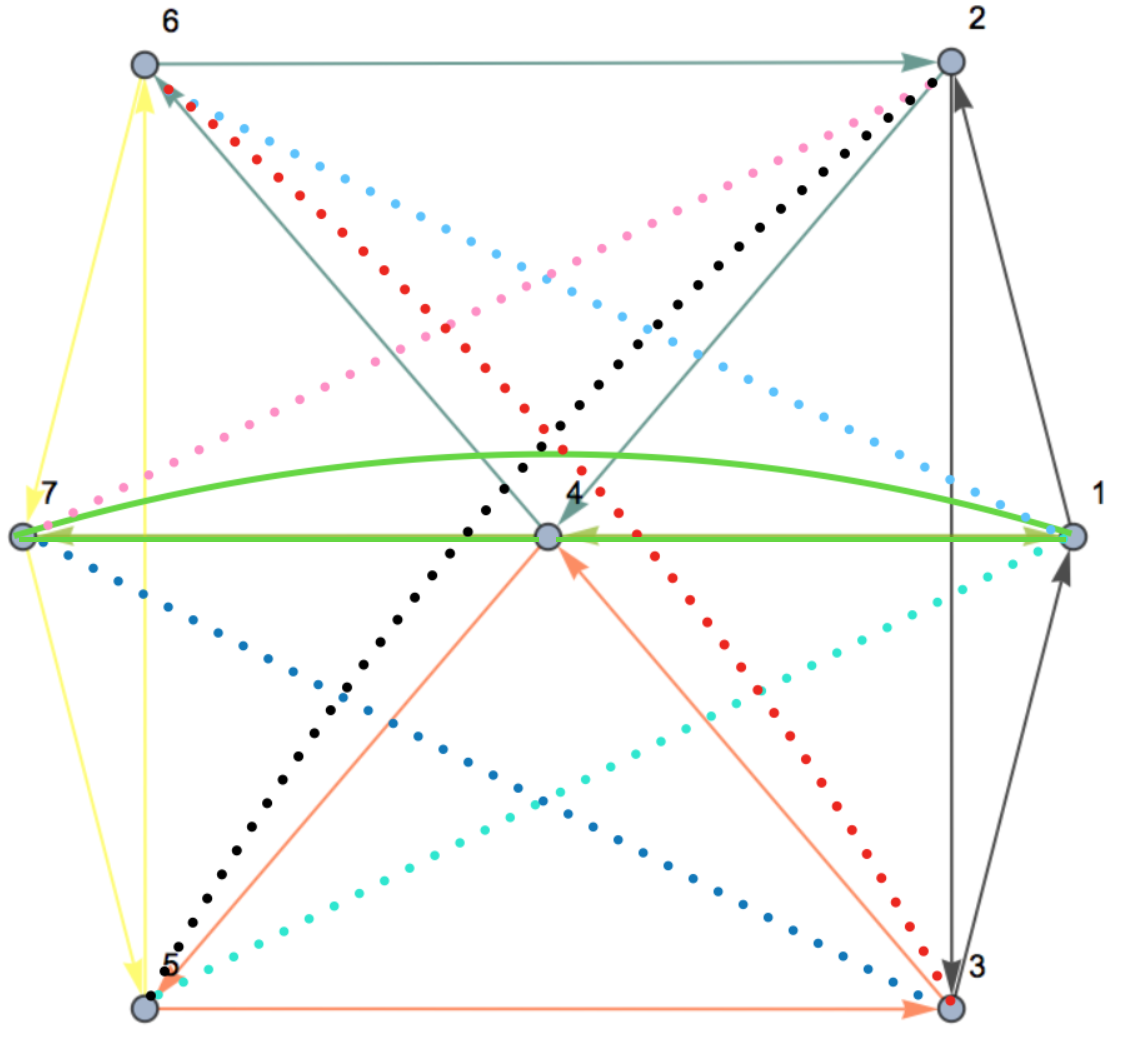}}
\caption{ Diagrammatic representation for LS functions}
\end{figure}

For \eqref{guebiw}, it is convenient here to use a diagrammatic representation, see Figure \ref{6ptir}, where each triangle of a distinct color represents a triplet.
We first make the diagram a complete graph for six points: since the four triangles have 12 edges, we need to add $15-12=3$ more (dashed) lines, $(14), (25), (36)$. Our construction is that we first choose $s_{14}, s_{25}, s_{36}$ to be constant, since the sum of all $n(n{-}1)/2=15$ $s_{i,j}$'s equals zero, we have
\be
s_{14}+s_{25}+s_{36}+s_{123}+s_{345}+s_{561}+s_{246}=0\,.
\ee
By choosing any three of the four $s_{i,j,k}$'s to be constant, the last one must be as well. Therefore, we can choose the hyperplane to be
\be\label{6ptcon}
H_6^{\rm irr}=\{
s_{1 4}, s_{2 5}, s_{3 6}, s_{1 2 3}, s_{3 4 5}, s_{5 6 1}~{\rm are~constants}\}\,,
\ee
which implies $s_{2 4 6}$ is also a constant. To see this indeed gives the desirable results, we study both the scattering and worldsheet forms. Note that there are in total $15$ $s_{i,j}$'s and $10$ $s_{i,j,k}$'s, thus apart from the 7 constants, we have 18 Mandelstam variables left. Exactly 32 cubic tree graphs can be built from these 18 poles that are not constant; out of the $7!!=105$ cubic trees, precisely these 32 have non-zero pullbacks to $H_6^{\rm irr.}$, and the coefficients are nothing but $\pm 1$. On the other hand, the pullback of scattering equations to  $H_6^{\rm irr}$ gives the Jacobian  with respect to {\it e.g.} $s_{12},s_{34},s_{56}$, and we have a beautiful determinant formula:
\ba\label{6ptjac}
J_{H_6^{\rm irr}}=(156)^{-1}
{\rm det}\left|
\begin{array}{ccc}
\<126\> &\<624\> & \<423\> \\
\<531\> &\<435\> & \<132\> \\
\<645\> &\<342\> & \<246\>
\end{array}
\right| \,,
\ea
where the abbreviation $\<ijk\>:=\frac{1}{\s_{ji}}-\frac 1 {\s_{jk}} $  is the inverse-soft factor  appearing in \eqref{ISF}. Of course we find that it is identical to the LS function, \eqref{guebiw}.

\begin{figure}[!htb]
\centering
\includegraphics[scale=1]{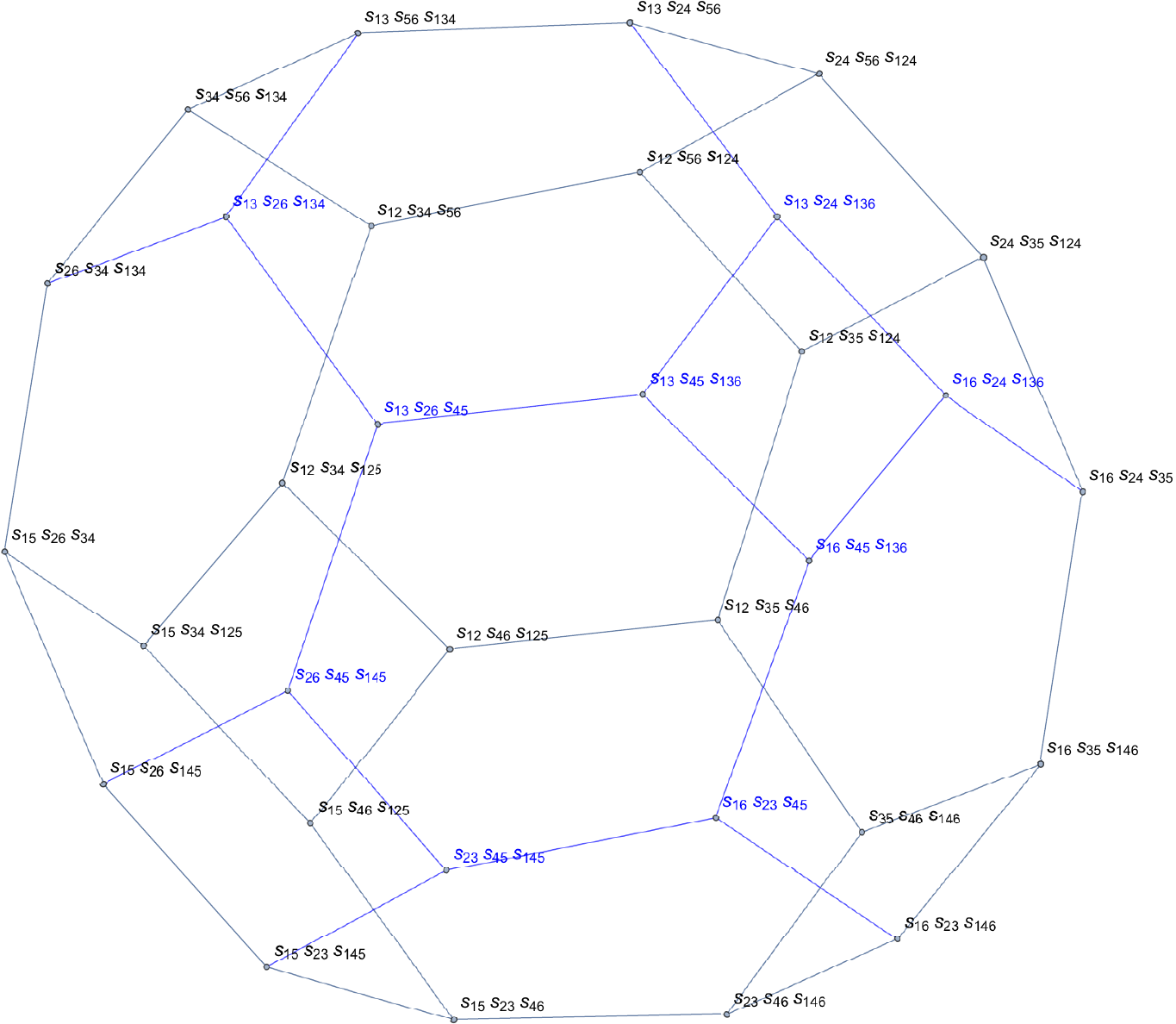}
\caption{Combinatoric polytope  from $\mathbf{LS}(\{(1,2,3),(3,4,5),(5,6,1),(2,4,6)\})$ \label{32irre}}
\end{figure}

Again similar to Cayley cases of \cite{Arkani-Hamed:2017mur} and \cite{Gao:2017dek}, we can construct the combinatorial polytope from the scattering form, corresponding to this LS function, see Figure \ref{32irre}

Let's study the irreducible example of $n=7$, see Figure \ref{7ptir}.
Again we first make it a complete graph by adding 6 dashed lines, corresponding to $s_{15},s_{16},s_{25},s_{27},s_{36},s_{37}$. The 5 triangles corresponds to $s_{123},s_{345},s_{567},s_{147},s_{246}$, and again the sum of all these variables equals zero. We can choose the hyperplane to be
\ba
H_7^{\rm irr}
=\{
s_{15},s_{16},s_{25},s_{27},s_{36},s_{37},s_{123},s_{345},s_{567},s_{714} ~{\rm are~constants}
\}\,,
\ea
which implies $s_{246}$ is also a constant. Out of the $9!!=945$ $W_g$'s, we find exactly 132 cubic trees have non-zero pullbacks (with coefficients $\pm 1$) to $H_7^{\rm irr.}$.
The Jacobian of scattering equations with respect to {\it e.g.} $s_{12},s_{34},s_{56},s_{47}$, again equals to the LS function:
\ba
 \frac{
{\rm det}\left|
\begin{array}{cccc}
\<126\> &\<624\> & \<426\>& \<623\> \\
\<531\> &\<435\> & 0 & \<132\> \\
\<645\> &\<342\> & \<241\> & \<746\>\\
\<354\>& \<753\>& \<657\>& 0
\end{array}
\right|}
{(167)} =
\frac{ {\rm det}^2\left|
\begin{array}{ccccc}
\s_{23} &\s_{31} &\s_{12}& 0 & 0\\
 0 & 0 & \s_{45}&\s_{53}& \s_{34}\\
\s_{56} & 0 & 0 & 0 & \s_{61}\\
0 &\s_{46}& 0 &\s_{62} & 0\\
\s_{47}& 0 & 0 &\s_{71} &0
\end{array}
\right| }
{\s_{67}^2(123)(345)(561)(246)(147)}\,.
\ea

Now we are ready to present the general proposition for the hyperplane that corresponds to any irreducible LS cases:
\begin{prop*}
One can choose the triplets for any irreducible LS $(\{i,j,k\})$ such that each label appears in more than one triplet, and each pair of labels appears at most in one triplet. There are exactly $3(n{-}2)$ edges from the $n{-}2$ triangles, thus we need $n(n{-}1)/2-3(n{-}2)=(n{-}3)(n{-}4)/2$ dashed lines $\{(a,b)\}$ to make a complete graph. After setting the corresponding $s_{a,b}$'s to constant, we further choose {\it any} $n{-}3$ of the $n{-}2$ $s_{i,j,k}$'s to be constant (which implies the last one is also constant). This is our proposal for the $d\log$ hyperplane corresponding to any irreducible LS function.
\end{prop*}

Finally, to get a general LS hyperplane, we first find the hyperplane for its irreducible part, then proceed by IS construction \eqref{IScondition}.
For example, the subspace of the IS-reducible function $\mathbf{LS}(\{(1,2,3),(3,4,5),(5,6,1)$ $,(2,4,6),\underline{(2,4,7)}\}) $ is
\ba
H_7=\{
s_{14}, s_{25}, s_{36}, s_{123}, s_{345}, s_{561}, \underline{s_{17},s_{37},s_{57},s_{67}}  \text{~are~constants}\}\,.
\ea

Unlike the IS-constructible LS functions, it is not obvious how to directly prove our proposition, {\it i.e.} to show that our subspaces yield the correct $d\log$ forms for general LS. In the following,  
we use a different strategy: given a LS function or any $d\log$ form on worldsheet, we will present a algorithem for finding a class of subspaces that yield such a LS function/$d\log$ form.  We will see the answer to this question also further provides some insights into the relation between subspaces, leading singulaties and general $d\log$ forms.


\paragraph{Constructing subspaces for general dlog forms}

By the pullback to some hyperplane $H$, we can obtain a $(n-3)\times n$ matrix from the scattering equations. However, in the context of leading function, the starting point is a $(n-2)\times n$ matrix. To prove that they have the same reduced determinant, it is better to state this proposition in terms of differential form. Since the leading singularity functions are originally defined in the $\lambda$-space, we introduce here a $(2n{-}4)$-differential form for any LS function $\mathcal{LS}(T_n)$
\begin{equation}
\Omega_{\mathrm{LS}}(T_n) = \mathcal{LS}(T_n)\frac{d^{2n}\lambda}{\mathrm{GL}(2)} \:, \label{lsform}
\end{equation}
and it can be easily rewritten as 
\begin{equation}
\Omega_{\mathrm{LS}}(T_n)= \bigwedge_{\tau\in T_n} 
d\log\frac{\langle\tau_{1}\tau_{2}\rangle}{\langle\tau_{3}\tau_{1}\rangle}\wedge d\log\frac{\langle\tau_{2}\tau_{3}\rangle}{\langle\tau_{3}\tau_{1}\rangle} \:, \label{lsform2}
\end{equation}
as shown in \cite{Arkani-Hamed:2014bca}. Now if we make a variable substitution $\lambda_{i}=t_{i}(1,\sigma_{i})$ like above and decompose the $\mathrm{GL}(2)$ redundancy to $\mathrm{SL}(2)$ for $\sigma$'s and $\mathrm{GL}(1)$ for $t$'s, then this $(2n-4)$-form will decompose into a $(n-3)$-form for $\sigma$'s and a $(n-1)$-form for $t$'s. This decomposition is quite trivial in (\ref{lsform}). It is simply
\begin{equation*}
\mathcal{LS}(T_n)\frac{d^{2n}\lambda}{\mathrm{GL}(2)}=\mathbf{LS}(T_n)\frac{d^n\sigma}{\mathrm{SL}(2)} \frac{d^{n}\log t}{\mathrm{GL}(1)} \:.
\end{equation*}
However such decomposition will lead to a nontrivial $(n-3)\times n$ matrix representation for LS function if we do it in (\ref{lsform2}).

This decomposition is quite straightforward even in (\ref{lsform2}). Since there is a trivial identity $d\log x\wedge d\log y=d\log (xy)\wedge d\log y= d\log (x/y)\wedge d\log y$, the $t$ factor in the first $d\log$ factor can be canceled by multiplying or dividing the terms in the other $d\log$ factors. This procedure can be proceeded until $n-3$ such $d\log$ appear, and the differential form consisting of these $n-3$ $d\log$'s is desired since it is already the top form on the $\sigma$-space. There is an example for the triple set $\{(1,2,5),(1,3,5),(1,4,5)\}$:
\begin{align}
&\quad d\log\frac{\langle 1\,2 \rangle}{\langle 5\,1\rangle}d\log\frac{\langle 2\,5 \rangle}{\langle 5\,1\rangle}d\log\frac{\langle 1\,3 \rangle}{\langle 5\,1\rangle}d\log\frac{\langle 3\,5 \rangle}{\langle 5\,1\rangle}d\log\frac{\langle 1\,4 \rangle}{\langle 5\,1\rangle}d\log\frac{\langle 4\,5 \rangle}{\langle 5\,1\rangle} \nonumber \\
&=d\log\frac{t_{2}}{t_{5}}\frac{\sigma_{12}}{\sigma_{51}}d\log\frac{t_{2}}{t_{1}}\frac{\sigma_{25}}{\sigma_{51}}d\log\frac{t_{3}}{t_{5}}\frac{\sigma_{13}}{\sigma_{51}}d\log\frac{t_{3}}{t_{1}}\frac{\sigma_{35}}{\sigma_{51}}d\log\frac{t_{4}}{t_{5}}\frac{\sigma_{14}}{\sigma_{51}}d\log\frac{t_{4}}{t_{1}}\frac{\sigma_{45}}{\sigma_{51}} \nonumber \\
&=d\log\frac{\sigma_{12}\sigma_{35}}{\sigma_{25}\sigma_{13}}d\log\frac{\sigma_{13}\sigma_{45}}{\sigma_{14}\sigma_{35}}d\log\frac{t_{1}}{t_{2}}\frac{\sigma_{51}}{\sigma_{25}}d\log\frac{t_{3}}{t_{1}}\frac{\sigma_{35}}{\sigma_{51}}d\log\frac{t_{4}}{t_{5}}\frac{\sigma_{14}}{\sigma_{51}}d\log\frac{t_{4}}{t_{1}}\frac{\sigma_{45}}{\sigma_{51}} \nonumber \\
&=d\log\frac{\sigma_{12}\sigma_{35}}{\sigma_{25}\sigma_{13}}d\log\frac{\sigma_{13}\sigma_{45}}{\sigma_{14}\sigma_{35}}d\log\frac{t_{1}}{t_{2}}d\log\frac{t_{3}}{t_{1}}d\log\frac{t_{4}}{t_{5}}d\log\frac{t_{4}}{t_{1}}\:,
\end{align}
where we have omitted the wedge product symbols for saving space. It is obvious that the variables of the $(n-3)$-form on the $\sigma$-space are some cross-ratios of $\sigma$'s. Actually, they are face variables of on-shell digram (see \cite{ArkaniHamed:2012nw}). In the following we will denote these variables as $f$'s and the corresponding $d\log$ factors as $d\log f$'s.

Thus we obtain another $(n-3)\times n$ matrix from those face variable by taking the partial derivative with respect to $\sigma_{i}$. It is obvious that the reduced determinant of this matrix gives the corresponding LS function. For example, three $f$'s for the triple set $\{(1,2,3),(3,4,5),(5,6,1),(2,4,6)\})$ can be chosen as $f_{1}=(\sigma_{13}\sigma_{26}\sigma_{45})/(\sigma_{12}\sigma_{46}\sigma_{35})$, $ f_{2}=(\sigma_{15}\sigma_{34}\sigma_{26})/(\sigma_{16}\sigma_{35}\sigma_{24})$ and $ f_{3}=(\sigma_{13}\sigma_{24}\sigma_{56})/(\sigma_{15}\sigma_{23}\sigma_{46})$, and the corresponding derivative matrix is 
\begin{equation}
\left(\frac{\partial \log f_{\alpha}}{\partial \sigma_{a}}\right)_{\alpha a} =
\begin{pmatrix}
\langle 213\rangle & \langle 126\rangle & \langle 531\rangle & \langle645\rangle & \langle 354\rangle & \langle 462\rangle \\
\langle516\rangle & \langle624\rangle & \langle435\rangle & \langle342\rangle & \langle153\rangle & \langle261\rangle \\
\langle315\rangle & \langle423\rangle & \langle132\rangle & \langle246\rangle & \langle651\rangle & \langle564\rangle 
\end{pmatrix} \:. \label{2.1}
\end{equation}
Remarkably, this matrix is the same as the matrix 
\begin{equation}
\left( \frac{\partial E_{a}}{\partial X_{\alpha}}\right)\bigg\rvert_{H}\,,
\end{equation}
where $X_{i}=\{s_{12},s_{34},s_{56}\}$ and the subspace $H$ is (\ref{6ptcon}). Then the reduced determinant of course gives the desired LS function. 
For the general case, we know the scattering equation $E_{a}$ comes from the derivative of Koba-Nielsen factor  $\mathcal{I}_{n}=\prod_{i<j}^{n}\sigma_{ij}^{s_{ij}}$ with respect to $\sigma_{a}$,
\[
E_{a}= \frac{1}{\mathcal{I}_{n}}\frac{\partial \mathcal{I}_{n}}{\partial \sigma_{a}}=\frac{\partial \log \mathcal{I}_{n}}{\partial \sigma_{a}}\,,
\]
then we have 
\begin{equation}
\left( \frac{\partial E_{a}}{\partial X_{\alpha}}\right)\bigg\rvert_{H} = \left( \frac{\partial^{2} \log \mathcal{I}_{n}}{\partial X_{\alpha}\partial \sigma_{a}}\right)\bigg\rvert_{H}\,.
\end{equation}
Thus the matrix $\partial \log f/ \partial \sigma_{a}$ is equal to the matrix $(\partial E_{a}/\partial X_{\alpha})\vert_{H}$ once we have 
\begin{equation}
\log f_{\alpha} = \frac{\partial \log\mathcal{I}_{n}}{\partial X_{\alpha}}\bigg\vert_{H} \:. \label{2.4}
\end{equation}
Now our question becomes: \emph{given a set of face variables $\{f_{\alpha}\}$, how to find a subspace such that (\ref{2.4}) holds?}

\, it is convenient to rewrite the factor $\log \mathcal{I}_{n}$ as a inner product
\begin{equation}
\log \mathcal{I}_{n} =\sum_{I=1}^{n(n-3)/2} Y^{I} \log U_{I}\,,
\end{equation}
where $Y^{I}$ form a basis of momentum space, and $u_{I}$ are some cross-ratios of $\sigma$'s such that the logarithm of any cross-ratio $V$ can be written as a linear combination of $\log U$'s with numerical coefficients. Then it is always able to find a linear transformation $\Lambda\indices{_I^J}$ such that $\log U_{I}=\sum\Lambda\indices{_I^J}\log V_{J}$ 
with $V_I=f_I$ for $I=1,2,\cdots,n-3$
. Then
\begin{equation}
\log \mathcal{I}_{n} = \sum_{I} Y^{I} \log U_{I} = \sum_{J}X^{J} \log V_{J}\,,
\end{equation}
where
\[
X^{J}=\sum_{I} Y^{I}\Lambda\indices{_I^J} \:.
\]
Remarkably, if we take the first $n-3$ of $X$'s as the basis of subspace and the others as the constant conditions, the (\ref{2.4}) trivially holds.

In the following we take the planar variables $s_{i,i+1,\cdots,j-1}$ as the basis of momentum space, and use the above 6-pt irreducible leading singularity as a example to illustrate how this formalism works. It is easy to find
\begin{align}
d\log\frac{\sigma_{13}\sigma_{26}\sigma_{45}}{\sigma_{12}\sigma_{46}\sigma_{35}} &= -d\log u_{13}+d\log u_{46} \:,\\
d\log\frac{\sigma_{15}\sigma_{34}\sigma_{26}}{\sigma_{16}\sigma_{35}\sigma_{24}}&= d\log u_{35} - d\log u_{26}  \:, \\
d\log\frac{\sigma_{13}\sigma_{24}\sigma_{56}}{\sigma_{15}\sigma_{23}\sigma_{46}}&= -d\log u_{24}+d\log u_{15} \:,
\end{align}
where
\[
u_{ij} := \frac{\sigma_{i-1,j}\sigma_{i,j-1}}{\sigma_{i-1,j-1}\sigma_{i,j}} \:,
\]
then we obtain the first three rows of $\Lambda^{-1}$ and denote it as $A$
\begin{equation}
A= \begin{pmatrix}
-1 & 0 & 0 & 0 & 0 & 0 & 0 & 0 & 1 \\
 0 & 0 & 0 & 0 & 0 & -1 & 1 & 0 & 0 \\
 0 & 0 & 1 & -1 & 0 & 0 & 0 & 0 & 0
\end{pmatrix}\,,
\end{equation}
where we have chosen $U_{I}$ to be the column vector $(u_{13},u_{14},u_{15},u_{24},u_{25},u_{26},u_{35},u_{36},u_{46})$. Actually, what we want is the last 6 columns of $\Lambda$ , which give the constant condition, and we denote the matrix consisting of those 6 columns as $A^{\bot}$ since it is orthogonal to $A$. A straightforward calculation then gives
\begin{equation}
(A^{\bot})^{\mathrm{T}}=\begin{pmatrix}
 1 & 0 & 0 & 0 & 0 & 0 & 0 & 0 & 1 \\
 0 & 0 & 0 & 0 & 0 & 0 & 0 & 1 & 0 \\
 0 & 0 & 0 & 0 & 0 & 1 & 1 & 0 & 0 \\
 0 & 0 & 0 & 0 & 1 & 0 & 0 & 0 & 0 \\
 0 & 0 & 1 & 1 & 0 & 0 & 0 & 0 & 0 \\
 0 & 1 & 0 & 0 & 0 & 0 & 0 & 0 & 0 \\
\end{pmatrix} \:.
\end{equation}
This gives a constant condition for this LS function:
\begin{align*}
s_{12}+s_{45} &= \text{const}. \qquad
s_{345} = \text{const}. \\
s_{16}+s_{34} &= \text{const}. \qquad
s_{234} = \text{const}. \\
s_{56}+s_{23} &= \text{const}. \qquad
s_{123} =\text{const}.
\end{align*}
which is equivalent to the previous constant condition since $s_{12}+s_{45}=s_{345}+s_{123}+s_{36}$, $s_{16}+s_{34}=s_{234}+s_{345}+s_{25}$, and $s_{56}+s_{23}=s_{123}+s_{234}+s_{14}$, as it should be. 

To summarize, we first find the expansion coefficients of $f_{\alpha}$ on the basis $U_{I}$, that is the $\frac{n(n-3)}{2}\times (n-3)$ matrix $A$, then its orthogonal complement $A^{\bot}$ gives the desired constant condition. We emphasise here that this applies beyond the case of leading singularities: \emph{any differential form on worldsheet can be obtained from a subspace as long as it can be written as a single term $d^{n-3}\log f$, and vice-versa.} It would be highly desriable to use this general construction for providing a proof for the propostion for any LS function.

\section{Outlook}

In this paper we have developed a subspace-based construction for projective scattering forms in ${\cal K}_n$ and the corresponding worldsheet forms in ${\cal M}_{0,n}$, which are related by pushforward by summing over solutions of scattering equations. It is natural to rewrite scattering equations as a map from ${\cal M}_{0,n}$ to any such subspace, where the Jacobian of transformation gives the wordsheet form. As the simplest and most elegant examples, we constructed subspaces, forms and the rewriting for Cayley cases, as well as polytopes in ${\cal K}_n$ and ${\cal M}_{0,n}$. We propose that it can be generalized to all cases corresponding to LS functions, both for IS constructible ones and beyond. There are lots of open questions raised by our initial exploration in these directions, including the following. 

\paragraph{Combinatorics and geometries of LS cases}

For non-Cayley LS cases, we have focused exclusively on the forms but not the geometries of polytopes in kinematic space. For IS constructible cases, we expect to construct convex polytopes in a way very similar to Cayley polytopes. Again we start with the cone where all poles are positive, and it should be possible to choose the constants defining $H$ to be negative or positive as appropriate; the intersection of the cone with $H$ must then yield a convex polytope, whose canonical form gives $\Omega_H$. We have checked this explicitly for the n=6 IS-constructible example in Figure \ref{22isf}.

On the other hand, it is still an open question to construct the convex polytope for general leading singularity cases. Already for the n=6 irreducible case in Figure \ref{32irre}, we see that the 18 poles add up to zero thus the same method cannot be applied. It seems we need new ideas to systematically construct such polytopes for general LS cases. Nevertheless, we do have strong evidences that at least combinatorially they are simple polytopes:
\begin{conj*}
All combinatoric polytopes obtained from LS functions are simple polytopes.
\end{conj*}
Obviously all combinatoric polytopes obtained from Cayley functions are simple polytopes by construction. 
Besides, according to our \textbf{claim} (I), we expect the $n$-pt polytope obtained form IS construction to be a simple polytopes if the $(n{-}1)$-pt one is. It would be interesting to prove this directly using our recursive construction.

We do not have a good idea how to prove even combinatorially irreducible LS functions correspond to simple polytopes. These are highly non-trivial objects, which almost certainly go beyond the scope of generalized permutohedra~\cite{postnikov2009permutohedra}. However, we have computed the $f$-vectors for these objects up to $n=10$, which include hundreds of types as shown in table~\ref{tab:fvector}. It is a very non-trivial test that all these $f$-vectors satisfy the Dehn-Sommerville equations ~\cite{stanley1986enumerative}, which strongly suggests that all of them are simple polytopes. In particular, we have seen that these are all Eulerian poset, {\it i.e.} each case has equal total number of odd/even-dimensional faces (if we count $f_{-1}=1$ and $f_{n-3}=1$). This is certainly not guaranteed, since we have found numerous objects from dlog subspaces beyond LS cases that are {\it not} Eulerian poset.

\paragraph{Worldsheet forms and regions for LS cases}

Apart from scattering forms and polytopes in kinematic space, the forms and regions in the moduli space are of great interests as well. The most important open question is to prove our {\bf{proposition}} about subspaces for LS cases, which is equivalent to (\ref{auwofhls}): given the {\bf{theorem}}, it suffices to show that the Jacobian of scattering equations with respect to the subspace equals the LS function, which would recursively prove \eqref{auwofhls}. As shown in~\cite{Cachazo:2017vkf}, on any co-dimensional one boundaries, LS function factories into two such functions. We believe that it is possible to show the same factorization for the Jacobian, which would recursively prove our \textbf{proposition}. Furthermore, the formula \eqref{auwofhls} for leading singularities were computed from MHV on-shell diagrams in ${\cal N}=4$ SYM, and it would be highly desirable to see if our determinant formula has any meaning in the context of SYM. Our construction for subspace is a good starting point to explore these questions. 

It would be also very interesting to study the worldsheet regions for general leading singularities. As shown in~\cite{Arkani-Hamed:2017mur}, we can choose any such region as the union of associahedra of different orderings, \eqref{cayre}. However, there are numerous choices of regions that gives the same LS function (up to a possible sign). We can easily see this since {\it e.g.} two different regions can add up to a region with vanishing canonical form, which can be interpreted as algebraic identities among Parke-Taylor factors. It would be nice to understand better all these regions, as well as the map via scattering equations.  From a wider perspective, just as the wordsheet associahedron ${\cal M}_{0,n}^+$ is the space from $G_{+}(2,n)$ modding out torus action, it would also be interesting to understand these regions as regions of $G(2,n)$, which correspond to MHV on-shell diagrams, modding out torus action. We hope to address these open questions in the future. 

\paragraph{More general forms and subspaces}

We have constructed explicitly forms and subspaces corresponding to all MHV leading singularities, but the method certainly extends to much more general cases, with several new directions to be explored. It is still an important open question to classify all subspaces that correspond to $d\log$ forms. In addition to search them from subspaces, one can also study pushforward of $d\log$ forms in the moduli space, {\it e.g.} the canonical form of general regions in ${\cal M}_{0,n}$. However, these more general $d\log$ forms are not as simple as the leading singularity cases. For example, already for $n=6$ we find several cases that do not correspond to simple polytopes but have the topology of a torus. It would be very interesting to study these more exotic objects. Another natural question is to see how our construction is related to other polytopes in the literature, such as generalized permutohedra~\cite{postnikov2009permutohedra} (beyond the Cayley cases) and cluster associahedra~\cite{FredericChapoton}. Very recently, there have been explorations of mathematical structures related to scattering forms, worldsheet forms and the geometries (see for example~\cite{Mizera:2017rqa,Early:2017tob,Salvatori:2018fjp,delaCruz:2017zqr,Torres:2017zzd,Frost:2018djd,Early:2018zuw}). It would be interesting to see how some of them fit in our picture as well. 

Another direction is to explore how our picture can be useful for the study of scattering forms and amplitudes in known theories. Given the recent progress for loop-level generalizations of CHY and ambitwistor strings, it is natural to study how to extend the associahedron and other geometries to one loop~\cite{Casali:2014hfa,Geyer:2015bja,He:2015yua,Geyer:2015jch,Cachazo:2015aol,Geyer:2016wjx,He:2016mzd,He:2017spx,Gomez:2016cqb}. Even for the Cayley cases, though individually these forms have not been interpreted as ``partial amplitudes" like for planar scattering forms, they are special combinations of cubic Feynman diagrams with nice properties such as factorizations. We have already seen their significance in the study of scattering forms in theories of physical interests, such as YM and NLSM. 

On the other hand, it would be extremely interesting if we could find subspaces to directly realize YM/NLSM forms, and it is not obvious that this is possible even for $n=5$. As shown in Appendix A, our {\bf{theorem}} actually generalizes to cases beyond subspaces: one can use any $d{-}(n{-}3)$ form (which generally does not correspond to any subspace) to obtain pullback of any $n{-}3$ form in kinematic space. This generalized way of pullback must work for YM/NLSM cases, and we expect it to have intriguing implications for color/kinematics duality and BCJ double copy. 
 
\section*{Acknowledgement} We are grateful to Nima Arkani-Hamed, Yuntao Bai for related collaborations and numerous stimulating discussions, and also Freddy Cachazo, Nick Early, Hugh Thomas for inspiring discussions. S.H. thanks the hospitality of IAS, Princeton where the work was finished. S.H.'s research is supported in part by the Thousand Young Talents program , the Key Research Program of Frontier Sciences of CAS under Grant No. QYZDB-SSW-SYS014 and Peng Huanwu center under Grant No. 11747601.

\appendix

\section{Proof of the main claim}\label{appa}
We prove~\eqref{pushforward} based on section~\ref{cayleyley}. Since both sides of~\eqref{pushforward} are decided locally by the tangent space of $H$, it suffices to consider the case where $H$ is a hyperplane described by $\frac{(n-2)(n-3)}{2}$ linear constraints, and it has global coordinates $X_1,...,X_{n-3}$. Choose $s_{a,b}$ with $2\le a<b\le n$ and $(a,b)\neq (2,n)$ as the $\frac{n(n-3)}{2}$ coordinates of $\mathcal{K}_n$. For any $(n-3)$ variables $z_1,z_2,...,z_{n-3}$ where each is a function of $s_{a,b}$'s, by the chain rule we have:
\begin{align}
\label{decdecdecpp}
	\biggl\lvert\frac{\partial z_i}{\partial X_j}\biggr\rvert_{H}=\sum_{\tilde{C}_n}\biggl\lvert\frac{\partial z_i}{\partial s{\scriptstyle(\tilde{C}_n)}_k}\biggr\rvert_{H(\tilde{C}_n)}\times\biggl\lvert\frac{\partial s{\scriptstyle(\tilde{C}_n)}_k}{\partial X_j}\biggr\rvert_{H}
\end{align}
where in the RHS the sum is over all ways to choose $(n-3)$ from the $s_{a,b}$'s mentioned above as the intermediate variables, namely, sum over all possible graph $\tilde{C}_n$ defined just like $C_n$ in section~\ref{cayleyley} but containing edge $(2,n)$. The $s{\scriptstyle(\tilde{C}_n)}_k$ with $k=1,..,(n-3)$ denotes the variables of a certain choice, or to say $s_{a,b}$ where $(a,b)$ is an edge in $C_n$ other than $(2,n)$. The hyperplane $H(\tilde{C}_n)$ is defined that those $s_{a,b}$'s not correspond to edges in $\tilde{C}_n$ are all constants, so it has been discussed in section~\ref{cayleyley}. Substitute $z_i$ in~\eqref{decdecdecpp} with $s_i^{(g)}$ and $E_i$ in~\eqref{ngngng} and~\eqref{Jac} respectively, we find that~\eqref{pushforward} is equivalent to:
\begin{align}
\label{decfacdecfac}
\sum_{\tilde{C}_n}\sum_{\rm sol.}~\omega^{(n{-}3)}_{H(\tilde{C}_n)}(s{\scriptstyle(\tilde{C}_n)}, \sigma)\times\biggl\lvert\frac{\partial s{\scriptstyle(\tilde{C}_n)}_k}{\partial X_j}\biggr\rvert_{H}=\sum_{\tilde{C}_n}\Omega_{H(\tilde{C}_n)}^{(n{-}3)}(s{\scriptstyle(\tilde{C}_n)}, s)\times\biggl\lvert\frac{\partial s{\scriptstyle(\tilde{C}_n)}_k}{\partial X_j}\biggr\rvert_{H}
\end{align}
In section~\ref{cayleyley} we have verified $\sum_{\rm sol.}~\omega^{(n{-}3)}_{H(\tilde{C}_n)}(s{\scriptstyle(\tilde{C}_n)}, \sigma)=\Omega_{H(\tilde{C}_n)}^{(n{-}3)}(s{\scriptstyle(\tilde{C}_n)}, s)$ for all $\tilde{C}_n$, so~\eqref{decfacdecfac} follows immediately and the proof is completed.

\section{Proof of the Cayley polytope construction}\label{appb}
Here we prove that the object $P(C_n):=\Delta(C_n)\cap H(C_n)$ mentioned in~\eqref{pcpcpcpcpcaaa} is indeed a polytope and study its boundary structure in the meanwhile.

Obviously, if $P(C_n)$ is nonempty, then it must be a convex polytope.  Suppose $P(C_n)$ is nonempty for $n\le m$, let's study $P(C_{m+1})$ by considering its possible boundaries.

\begin{figure}
\centering
\begin{overpic}[width=0.9\linewidth]
{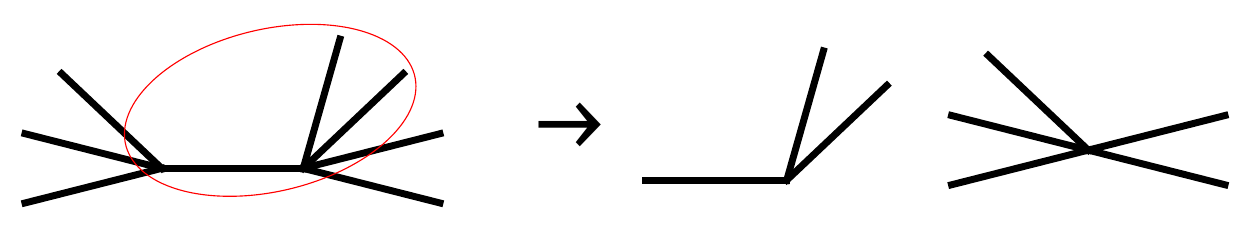}
\end{overpic}
\caption{Example for $A$ and $\tilde{A}$ from $C_n$}\label{illgrafac2}
\end{figure}

From~\eqref{rgrgrgrgraaa} we know that the boundaries of $P(C_{m+1})$ can only lie on the hyperplane $s_A=0$ for a certain nontrivial connected subgraph $A$ of $C_{m+1}$. With this fact, \eqref{rgrgrgrgraaa} can be written as
\begin{align}
\label{eqncdtndecmpdecmp}
\Delta(C_{m+1})\xrightarrow{s_A=0}&\{s_{AI}\ge 0\arrowvert AI\subsetneq A\}\cap\{s_{AO}\ge 0\arrowvert AO\supsetneq A\textrm{ or }AO\cap A=\emptyset\}\nonumber\\
\cap&\{s_{AS}\ge 0\arrowvert AS\cap A=B, B\notin\{A,AS,\emptyset\}\}
\end{align}
\begin{figure}
\centering
\begin{overpic}[width=0.7\linewidth]
{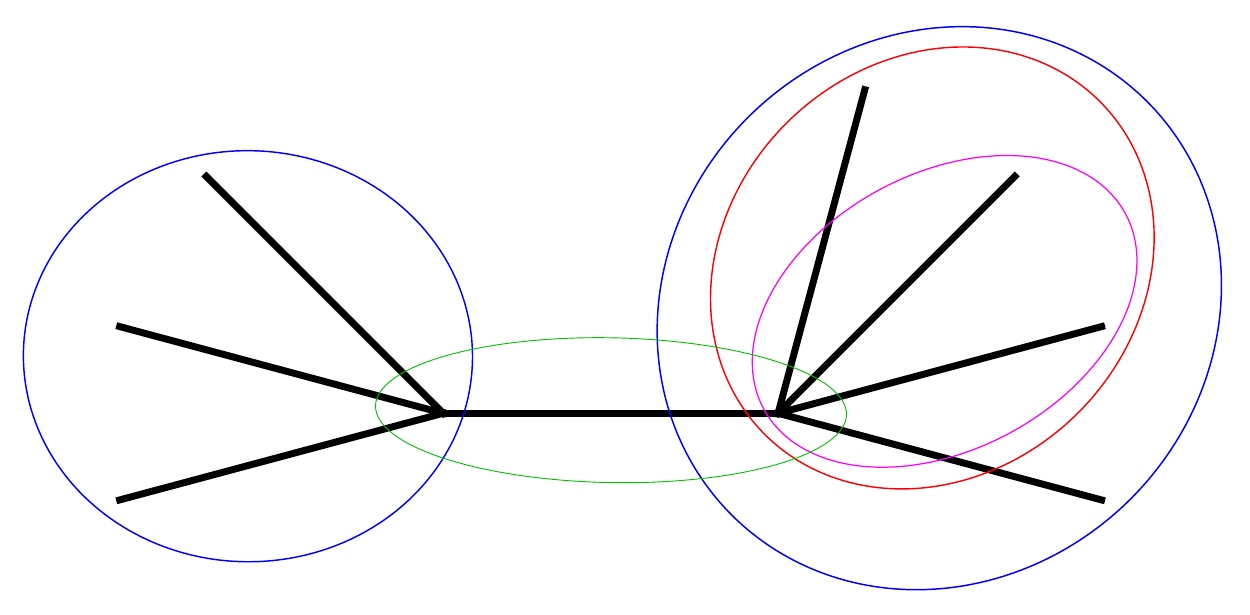}
\end{overpic}
\caption{Examples of \textcolor[rgb]{1,0,1}{$AI$}, \textcolor[rgb]{0,0,1}{$AO$}, \textcolor[rgb]{0,0.75,0}{$AS$} with given \textcolor[rgb]{1,0,0}{$A$} in $C_n$}\label{illgrafac}
\end{figure}
where $AI, AO, AS$ are all connected subgraphs (see Figure~\ref{illgrafac}). Since $C_{m+1}$ is a tree graph, $B$ and $AS\cup A$ are also connected. We have
\begin{align}
\label{eqnmstmstmst}
s_B+s_{AS\cup A}=s_A+s_{AS}+\sum_{\substack{i\in A\setminus AS \\j\in AS\setminus A}}s_{i,j} \:.
\end{align}
It's obvious that the $(i,j)$ pairs in the RHS of~\eqref{eqnmstmstmst} are not adjacent. Thus, the third term on RHS of~\eqref{eqnmstmstmst} is a negative constant on $H(C_{m+1})$, while $s_B$ and $s_{AS\cup A}$ remain nonnegative in the first and second set on RHS of~\eqref{eqncdtndecmpdecmp}, respectively. It implies $s_{AS}$ are automatically nonnegative with the support of $s_{AI}\geq0$ and $s_{AO}\geq 0$. In other words, we have
\begin{align}
\label{eqncdtndecmpft}
\Delta(C_{m+1})\xrightarrow{\{s_A=0\}\cap H(C_{m+1})}&\{s_{AI}\ge 0\arrowvert AI\subsetneq A\}\cap\{s_{AO}\ge 0\arrowvert AO\supsetneq A\textrm{ or }AO\cap A=\emptyset\} \:.
\end{align}
On the other hand, $H(C_{m+1})$ can be written as (the vertex pairs are all nonadjacent):
\begin{align}
\label{eqncdtdecmpft2}
H(C_{m+1})=\{s_{i,j}=-c_{i,j}\arrowvert\{i,j\}\subseteq A\}\cap\{s_{i,j}=-c_{i,j}\arrowvert\{i,j\}\nsubseteq A\} \:.
\end{align}
Since we already have the subgraph $A$, we can obtain another connected graph from $C_{n}$ by contracting $A$ to a single node, and denote this graph as $\tilde{A}$ (see Figure~\ref{illgrafac2}).)
Recall that when $s_A=0$, $s_{A\cup\{j\}}=\sum_{i\in A}s_{i,j}$ for $j\notin A$. So from~\eqref{eqncdtndecmpft} and~\eqref{eqncdtdecmpft2} we know that:
\begin{align}
\label{eqngeofac}
P(C_{m+1})\arrowvert_{s_A=0}=&(R(C_{m+1})\cap H(C_{m+1}))\arrowvert_{s_A=0}\nonumber\\
=&(\{s_{AI}\ge 0\arrowvert AI\subsetneq A\}\cap\{s_{i,j}=-c_{i,j}\arrowvert\{i,j\}\subseteq A\})\nonumber\\
&\cap(\{s_{AO}\ge 0\arrowvert AO\supsetneq A\textrm{ or }AO\cap A=\emptyset\}\cap\{s_{i,j}=-c_{i,j}\arrowvert\{i,j\}\nsubseteq A\})\nonumber\\
=&P(A)\cap P(\tilde{A})\cap\{\textrm{constants restraints irrelevant to $P(A)$ or $P(\tilde{A})$}\}
\end{align}
where the constants packed in $P(A)$ and $P(\tilde{A})$ are implicitly inherited from $H(C_{m+1})$. Thus $P(C_{m+1})$ is nonempty by the induction assumption. And from~\eqref{eqngeofac} we can infer the boundary structure of $P(C_n)$: it has a codimension-1 boundary at every $s_A=0$ with $A$ being $C_n$'s nontrivial connected graph, and the shape of this boundary is $P(A)\otimes P(\tilde{A})$. So each codimension-$k$ boundary of $P(C_n)$ must correspond to $k$ times of such successive ``factorize'' process, namely a set of $k$ compatible poles $s_{A_1},...,s_{A_k}$ where $A_i$ is connected in $C_n$. This completes the proof of the assertion in sec. \ref{cayleyley}.

\bibliographystyle{utphys}
\bibliography{bib}

\end{document}